\documentclass[fleqn]{article}
\usepackage{amssymb,latexsym}
\usepackage[dvips]{graphicx}
\DeclareFontFamily{OT1}{rsfs}{} \DeclareFontShape{OT1}{rsfs}{m}{n}{
<-7> rsfs5 <7-10> rsfs7 <10-> rsfs10}{}
\DeclareMathAlphabet{\mycal}{OT1}{rsfs}{m}{n}
\begin{document}
\title{``Peeling property'' for linearized gravity in null coordinates}
\author{Jacek Jezierski\thanks{Partially supported by a grant KBN Nr 2
P03A 047 15 and CNRS Orleans. E-mail: Jacek.Jezierski@fuw.edu.pl}\\
D\'epartement de Math\'ematiques, UMR 6083 du CNRS,\\
Universit\'e de Tours, Parc de Grandmont, F-37200 Tours, France\\
{\em on leave of absence:}\\
Department of Mathematical Methods in Physics, \\ University of Warsaw,
ul. Ho\.za 74, 00-682 Warsaw, Poland}
\date{PACS numbers: 11.10.Ef, 04.20.Ha, 11.30.Jj}
\maketitle


{\catcode `\@=11 \global\let\AddToReset=\@addtoreset}
\AddToReset{equation}{section}
\renewcommand{\theequation}{\thesection.\arabic{equation}}

\newtheorem{Lemma}{Lemma}
\AddToReset{Lemma}{section}
\newtheorem{Theorem}{Theorem}
\AddToReset{Theorem}{section}
\newfont{\msa}{msam10 scaled\magstep1}
\font\SYM=msbm10 
\def\Complex{\hbox{\SYM C}}
\def\Rationals{\hbox{\SYM Q}}
\def\Reals{\hbox{\SYM R}}
\def\Integers{\hbox{\SYM Z}}
\def\Naturals{\hbox{\SYM N}}

\newcommand{\dtwo}{\kolo{\Delta}}
\newcommand{\kolo}[1]{\vphantom{#1}\stackrel{\circ}{#1}\!\vphantom{#1}}
\newcommand{\ve}{\varepsilon}
\newcommand{\ol}{\overline}

\def\scri{{\mycal I}}
\def\scrip{\scri^{+}}%
\def\scrp{{\mycal I}^{+}}%
\def\Scri{\scri}

\newcommand{\ind}{\mathop {\rm ind}\nolimits }
\newcommand{\be}{\begin{equation}}
\newcommand{\ee}{\end{equation}}
\newcommand{\ssr}[1]{{\, \strut ^{w}}\! {#1}}
\newcommand{\ssd}[1]{{\, \strut ^{d}}\! {#1}}
\newcommand{\ssm}[1]{{\, \strut ^{m}}\! {#1}}
\newcommand{\rd}{{\rm d}} 
\newcommand{\E}[1]{{\rm e}^{#1}}
\newcommand{\ber}{\begin{eqnarray}}
\newcommand{\eer}{\end{eqnarray}}
\newcommand{\pl}{{\frac{1}{2}}}
\newcommand{\g}{\Gamma}
\newcommand{\ep}{\varepsilon}
\newcommand{\tr}{\ddagger}
\newcommand{\sk}{\sqrt{k}}
\newcommand{\skr}{\frac{\sqrt{k}}{r}}
\newcommand{\dwaskr}{\frac{2 \sqrt{k}}{r}}
\newcommand{\lpd}{{(\stackrel{\circ}{\Delta} + 2)}}
\newcommand{\gauge}{\mbox{\rm gage}}
\newcommand{\paragraf}[1]{\section{#1} \typeout{Paragraf \space
\thesection} }
\newcommand{\ten}[3]{{#1}^{#2}_{\ #3}}
\newcommand{\tenud}[3]{{#1}^{#2}{_{#3}}}
\newcommand{\tendu}[3]{{#1}_{#2}{^{#3}}}
\newcommand{\tenudu}[4]{{#1}^{#2}{_{#3}}{^{#4}}}
\newcommand{\tendud}[4]{{#1}_{#2}{^{#3}}{_{#4}}}
\newcommand{\wek}[2]{{#1}^{#2}}
\newcommand{\kowek}[2]{{#1}_{#2}}
\newcommand{\base}[2]{{{\partial}\over {\partial
{#1}^{#2}}}}
\newcommand{\diff}[2]{\frac{\partial{#1}}{\partial{#2}}}
\renewcommand{\varpi}{{\cal P}}
\renewcommand{\Upsilon}{\iota}
\begin{abstract}
A complete description of the linearized gravitational field on a
flat background is given in terms of gauge-independent quasilocal
quantities. This is an extension of the results from \cite{ISMC98}.
Asymptotic spherical quasilocal parameterization of the Weyl field and
its relation with Einstein equations is presented. The field
equations are equivalent to the wave equation.
A generalization for Schwarzschild background is developed
 and the axial part
of gravitational field is fully analyzed.
In the case of axial degree of freedom for linearized
gravitational field the corresponding generalization
of the d'Alembert operator is a Regge-Wheeler equation.
Finally, the asymptotics at null infinity is investigated and
strong peeling property for axial waves is proved.
\end{abstract}

\section{Introduction}

 We show that
seemingly complicated linearized Einstein equations on a
Schwarzschild background can be analyzed in terms of gauge
invariants. The obtained invariants decouple, in a natural way,
into axial and polar parts keeping symmetry with respect to the
 interchange of the null coordinates
$u$ and $v$. The invariant $\bf y$ describing axial
degrees of freedom, corresponding to $\Im \Psi_0$, fulfills
Regge-Wheeler equation because axial part of the corresponding
component of the Weyl field
is gauge-invariant\footnote{More precisely, some components
in axial part of the Weyl field are
gauge invariant but not all of them and they have to be
``corrected'' by metric terms (see Section \ref{stR}) to
become gauge independent.}. On the other hand
the polar part the Weyl field is not
gauge-invariant. However, all components of the Weyl field may be
``corrected'' in such a way that we obtain invariants which
substitute linearized Newman-Penrose scalars.

In \cite{ISMC98} we have shown how the gauge-invariant
quantities $\bf x$, $\bf y$ describing unconstrained degrees of freedom
of the gravitational field arise in a canonical formalism.
Here, we concentrate on their relations with linearized
Weyl tensor.
In the case of a flat background we present, in Theorem 1, an
explicit relation between linearized Weyl tensor and the invariants.
We continue this analysis
for the case of a Schwarzschild background and we show an analogous
relation but only for the axial degree of freedom described by $\bf y$.
Finally, we examine this result in view of the so-called
{\em peeling} property and we show that axial part of the linearized
gravitational field obeys strong peeling.

This paper is organized as follows: In the next Section, some
preliminary notions and results for the flat background are introduced.
Section 3 contains a generalization for the Schwarzschild background,
in particular, gauge invariants and Einstein equations for the axial part
of the gravitational field are presented.
In Section 4 we discuss the relation between invariants and linearized
Riemann tensor.
Section 5 is devoted to the investigation of the asymptotics at null
infinity for the solutions of Regge-Wheeler equation and its connection
with peeling property.
To clarify the exposition some of the technical results and proofs have been
shifted to the appendix.

\section{Description in null coordinates for flat background}
\label{ncfb}
We present in this section some standard results about linearized
gravitational field with nontrivial extensions not only in a notation but
also in the framework.

\subsection{Minkowski metric in null coordinates}
Let us consider the flat Minkowski metric of the following form in
spherical coordinates
\be \eta_{\mu\nu}\rd y^\mu \rd y^\nu = -\rd t^2 +\rd r^2 + r^2 (\rd
\theta^2 +\sin^2\theta \rd \phi^2) \, . \ee
The Minkowski space $M$  has a natural structure of a spherical
foliation around null infinity, more precisely, the neighbourhood of
$\Scri^+$ looks like $S^2 \times M_2$.
  We shall use several coordinates on $M_2$:
$t,r,\rho,v,u$. They are defined as follows
\[ \rho := r^{-1} \quad u:=t-r \quad  v:=t+r \, . \]
Let us fix the null coordinates $(u,v)$ together with the index $a$
corresponding to them.

 The coordinates on a sphere we denote $(x^A),(A=1,2)$, $(x^1=\theta,
x^2=\phi)$ and the round metric on a unit sphere by $\kolo\gamma_{AB}$
($\kolo\gamma_{AB}\rd x^A\rd x^B =\rd\theta^2 +\sin^2\theta \rd
\phi^2$). Let us also denote by $\dtwo$ the laplacian corresponding to
the metric $\kolo\gamma_{AB}$. Moreover, we use the symbol ``$||$'' for
the covariant derivative on $S^2$ with respect to the induced metric
$\eta_{AB}$.

For convenience we need also some more denotations:
$\rho =r^{-1}=\frac2{v-u}$, $\rho_{,a}=\rho^2 \varepsilon_a$
where $\varepsilon_u:=\frac12$, $\varepsilon_v:=-\frac12$,
$\eta^{ab}\varepsilon_a\varepsilon_b=1$. We define
$\varepsilon^a:=\eta^{ab}\varepsilon_b$
and one can check that
$\varepsilon^u=1$, $\varepsilon^v=-1$, $\eta_{ab}\varepsilon^a\varepsilon^b=1$.

The explicit formulae for the components of Minkowski metric can be
denoted as follows

\[ \eta_{AB}=\rho^{-2}\kolo\gamma_{AB} \, , \quad
 \eta_{ab}=-\frac12 |E_{ab}| \, , \quad
 \eta_{aA}=0 \]
 where $E_{uu}=0=E_{vv}$ and $E_{uv}=1=-E_{vu}$ and
 \[ \eta_{\mu\nu}\rd x^\mu \rd x^\nu =
 \eta_{ab}\rd x^a \rd x^b + \eta_{AB}\rd x^A \rd x^B = -\rd u \rd v + \rho^{-2}
 ( \rd\theta^2 +\sin^2\theta \rd \phi^2 ) \, .\]
 Similarily, the inverse metric has the following components
\[ \eta^{AB}=\rho^{2}\kolo\gamma^{AB} \, , \quad \eta^{ab}=-2 |E^{ab}| \,
 , \quad \eta^{aA}=0 \]
 where $E^{uu}=0=E^{vv}$ and $E^{uv}=1=-E^{vu}$.
We shall also need the derivatives
\[ \eta^{AB}{_{,a}}=2\rho\varepsilon_a\eta^{AB} \, , \quad
 \eta_{AB}{_{,a}}=-2\rho\varepsilon_a\eta_{AB} \]
and finally the nonvanishing Christoffel symbols are the following
\[ \Gamma^a{_{AB}}=\rho\varepsilon^a\eta_{AB} \, , \quad
 \Gamma^A{_{aB}}=-\rho\varepsilon_a\delta^A{_B}\, , \quad \Gamma^A{_{BC}} \]
where $\Gamma^A{_{BC}}$ are Christoffel symbols for the spherical covariant
derivative ``$||$'' on $S^2$.

\subsection{Riemann tensor in null coordinates}
The linearized Riemann tensor $R_{\mu\nu\lambda\delta}$ defined in an
obvious way in terms of the second derivatives of the linearized metric
$h_{\mu\nu}$ by the formula
\[
   2R_{\mu\nu\lambda\delta}:=
    h{_{\mu\delta;\nu\lambda}} -h{_{\nu\delta;\mu\lambda}} + h_{\nu\lambda;\mu\delta}
     - h{_{\mu\lambda;\nu\delta}}
 \]
has the following components in null coordinates:
\[
   2R_{abcd}= h{_{ad,bc}} -h{_{bd,ac}} + h_{bc,ad} - h{_{ac,bd}}
 \]

\[
   2R_{abcD}= h{_{aD,bc}} -h{_{bD,ac}} + h_{bc,aD} - h{_{ac,bD}} +
         \] \[
 +\rho\varepsilon_b \left( h_{aD,c} + h_{cD,a}- h_{ac,D} \right)
       - \rho \varepsilon_a \left( h_{bD,c} + h_{cD,b}- h_{bc,D} \right)
\]
\[
  2R_{AbCd}= h{_{dA||C,b}} +h{_{bC||A,d}} - h_{bd||AC} - h{_{AC,bd}} +
         \] \[
   +\rho\varepsilon_b \left( h_{dA||C} - h_{dC||A}- h_{AC,d} \right)
       + \rho \varepsilon_d \left( h_{bC||A} - h_{bA||C}- h_{AC,b} \right)+
\]
\[ +\rho\eta_{AC}\varepsilon^a \left( h_{bd,a}-h_{ad,b}-h_{ab,d}\right)
 -2\rho^2\varepsilon_b\varepsilon_d h_{AC}
\]

\[
  2R_{ABCd}= h{_{dA||BC}} +h{_{BC||A,d}} - h_{Bd||AC} - h{_{AC||B,d}}
   +2\rho \varepsilon_d \left( h_{BC||A} - h_{AC||B} \right)+ \]
  \[
 +\rho\eta_{BC} \varepsilon^a \left( h_{aA,d} - h_{dA,a}+ h_{ad,A}
 +2\rho\varepsilon_d h_{aA} \right)
 -\rho\eta_{AC}\varepsilon^a \left( h_{aB,d}-h_{dB,a}+h_{ad,B}
 +2\rho\varepsilon_d h_{aB} \right) \]

\[
  2R_{abCD}= h{_{aD||C,b}} -h{_{bD||C,a}} + h_{bC||D,a} - h{_{aC||D,b}} +
    \]
     \[
  +2\rho\varepsilon_b \left( h_{aD||C} - h_{aC||D} \right)
       +2 \rho \varepsilon_a \left( h_{bC||D} - h_{bD||C} \right)
\]

\[
   2R_{ABCD}= h{_{AD||BC}} +h{_{BC||AD}} - h_{BD||AC} - h{_{AC||BD}}
   +\]
    \[
   +\rho\eta_{AC}\varepsilon^a \left( h_{BD,a}-h_{aB||D}-h_{aD||B} \right)
 +\rho\eta_{BD}\varepsilon^a \left( h_{AC,a}-h_{aC||A}-h_{aA||C} \right)
 + \]
 \[
 -\rho\eta_{BC} \varepsilon^a \left(  h_{AD,a} - h_{aA||D} - h_{aD||A} \right)
 -\rho\eta_{AD}\varepsilon^a \left( h_{BC,a}-h_{aB||C}-h_{aC||B} \right)
 +\]
 \[
 +\rho^2 \left( h_{BD} \eta_{AC} +h_{AC}\eta_{BD}
-h_{AD}\eta_{BC}  -h_{BC}\eta_{AD} \right)
+2\rho^2\varepsilon^a\varepsilon^b h_{ab} \left( \eta_{AC}\eta_{BD}
-\eta_{BC}\eta_{AD} \right)
\]
We show in the sequel how the above formulae can be generalized
for the Schwarzschild background.

\subsection{Ricci tensor in null coordinates}
The linearized Ricci tensor $R_{\mu\nu}:=
\eta^{\delta\lambda}R_{\delta\mu\lambda\nu}$
 takes the following form in our coordinates
\[
       2R_{ab}= h^c{_{b,ac}}+h_a{^c}{_{,cb}} - h_{ab}{^{,c}}{_c} - h^c{_{c,ab}}
                + h_{aA,b}{^{||A}} +  h_{bA,a}{^{||A}} - h_{ab}{^{||A}}{_A}
        -H_{,ab}+ \] \[
+\rho\varepsilon_a H_{,b} +\rho\varepsilon_b H_{,a}
       + 2\rho \varepsilon^c \left( h_{ab,c} -h_{ac,b}- h_{bc,a} \right)
\]
\[
         2R_{aB}= h^b{_{B,ab}} -h_{aB}{^{,c}}{_c} +h_a{^c}{_{,cB}}
 - h^c{_{c,aB}} +h_{a}{^{A}}{_{||BA}} - h_{aB}{^{||A}}{_A}
+\chi_{B}{^{A}}{_{||A,a}} -\frac12 H_{||B,a} + \] \[
 +\rho\varepsilon_a \left( 2 h^b{_{B,b}} - h^b{_{b,B}} \right)
  -2\rho\varepsilon^b h_{bB,a} -2\rho^2\varepsilon_a\varepsilon^b h_{bB}
\]
\[
       2R_{AB}= \left( h^a{_{A||B}}+h^a{_{B||A}} \right)_{,a}
- h^a{_{a||AB}} - \chi_{AB}{^{,a}}{_{a}}-2\rho\varepsilon^a \chi_{AB,a}
  +\chi_A{^C}{_{||CB}} +\chi_B{^C}{_{||CA}}  + \]
\[ -\chi_{AB}{^{||C}}{_C}
+\eta_{AB} \left[ -\frac12 ( H^{||C}{_C} + H^{,a}{_a} )
   +2\rho\varepsilon^a (H_{,a} -h_a{^A}{_{||A}}) +\rho^2 (2\varepsilon^a
  \varepsilon^b h_{ab} -H )   \right]
\]
where $H:=\eta^{AB}h_{AB}$ and $\chi_{AB}:=h_{AB}-\frac12\eta_{AB} H$.
Some components of the linearized Ricci are also derived for
Schwarzschild background in Appendix \ref{apendix_rownania}.

\subsection{Gauge in null coordinates}
The gauge transformation $\xi_\mu$
\[ h_{\mu\nu}\longrightarrow h_{\mu\nu}+2\xi_{(\mu;\nu)}\]
splits in the following way
\[ h_{ab} \longrightarrow h_{ab}+\xi_{a,b}+\xi_{b,a}\]

\be\label{gaugep}
h_{aA}  \longrightarrow  h_{aA}+\xi_{a,A}+\xi_{A,a}+2\rho \varepsilon_a \xi_A
\ee

\[ h_{AB} \longrightarrow h_{AB}+\xi_{A||B}+\xi_{B||A}-2\rho
\eta_{AB}\varepsilon^a\xi_a \, . \]
There are also some useful formulae
\[\chi_{AB} \longrightarrow
\chi_{AB}+\xi_{A||B}+\xi_{B||A}-\eta_{AB}\xi^C{_{||C}}\]

\[\frac12 H  \longrightarrow \frac12 H +\xi^A{_{||A}}-2\rho\varepsilon^a\xi_a\]

\[ h_a{^A}  \longrightarrow h_a{^A} +\xi_a{^{||A}}+\xi^A{_{,a}}\]
which are straightforward consequences of the previous ones.

\subsection{Invariants and vacuum Einstein equations}
Let us introduce the following gauge invariant
quantities\footnote{We leave to the reader an
exercise to check that those quantities are gauge invariant (using (\ref{gaugep})),
 however, in the Appendix
\ref{apendix_invariants} we show this property in a more general case.}

\begin{eqnarray}\label{ya}
 {\bf y}_a &:=& (\dtwo
+2)h_{aA||B}\varepsilon^{AB}-(\rho^{-2}\chi_A{^C}{_{||CB}}
\varepsilon^{AB})_{,a}
\end{eqnarray}
\begin{eqnarray} \label{doy}
{\bf y} &:=& 2\rho^{-2}(h_{bB||A}\varepsilon^{AB})_{,a}E^{ab}
\end{eqnarray}
\begin{eqnarray}
 {\bf x} &:=& \rho^{-2}\chi^{AB}{_{||BA}}-\frac12 \dtwo H+\rho^{-1}\varepsilon^a
H_{,a}-H+2\varepsilon^a \varepsilon^bh_{ab}-2\rho^{-1}\varepsilon^ah_a{^A}{_{||A}}
\end{eqnarray}
\begin{eqnarray} {\bf x}_{ab} &:=& \dtwo (\dtwo +2) h_{ab} -(\dtwo +2)
\left[(\rho^{-2}h_a{^A}{_{||A}})_{,b} +(\rho^{-2}h_b{^A}{_{||A}})_{,a}
\right] \nonumber \\  & &
+\left[\rho^{-2}(\rho^{-2}\chi^{AB}{_{||AB}})_{,a} \right]_{,b}
  +\left[\rho^{-2}(\rho^{-2}\chi^{AB}{_{||AB}})_{,b} \right]_{,a}
\label{xab}
\end{eqnarray}
where $\varepsilon^{AB}$  is the
Levi-Civita skew-symmetric tensor on a sphere $\{ u={\rm const.},\,  v={\rm
const.} \}$ such that $\rho^{-2}\sin\theta\varepsilon^{12}=1$.

The axial invariants are not independent they are related as follows:
\[
  2\rho^{-2}{\bf y}_{b,a}E^{ab}
=(\dtwo +2){\bf y}
\]
which is a simple consequence of the definition (\ref{doy}).

If we assume that vacuum Einstein equations $R_{\mu\nu}=0$ are fulfilled, we obtain the
following equations for our invariants:

The axial part reads as
\be (\rho^{-2}{\bf y}^a)_{,a}=2\rho^{-4} \kolo{R}_A{^B}{_{||BD}}\varepsilon^{AD}=0
\ee

\be \label{yay} 2E^{ab}(\rho^{-2}{\bf y})_{,b} +\rho^{-2}{\bf y}^a=
-2\rho^{-4} {R}{^a}{_{B||D}}\varepsilon^{BD}=
 0 \ee
or takes another form in terms of the quantity ${\bf y}^a$
\[ [\rho^{-4}({\bf y}_{a,b}-{\bf y}_{b,a})]^{,b}+\rho^{-2}(\dtwo +2) {\bf
y}_a = -2\rho^{-4} (\dtwo +2){R}{_{aB||D}}\varepsilon^{BD}=0 \, . \]

The polar part takes the following form
\[ {\bf x}^{ab}{_{,ab}} - \rho^{2} \dtwo(\dtwo +2) {\bf x} =
4 ( \rho^{-4}\kolo{R}^{AB}{_{||BA}} )^{,a}{_{a}} +
\dtwo(\dtwo +2)\left( \eta^{ab}R_{ab} -\eta^{AB}R_{AB} \right) = 0 \]

 \be\label{etaxab} \eta^{ab} {\bf x}_{ab}=
 4\rho^{-4}\kolo{R}^{AB}{_{||BA}} = 0 \ee

 \be\label{xabx} {\bf x}_{ab} - 2(\rho^{-2}{\bf x})_{,ab}
  +\eta_{ab}(\rho^{-2}{\bf x})^{,c}{_c} =0 \, . \ee
The left-hand side of the last equation (\ref{xabx})
depends on all seven polar
components of the Ricci tensor $R_{ab}$, $R^{aB}{_{||B}}$,
$\eta^{AB}R_{AB}$, $\kolo{R}^{AB}{_{||BA}}$. Assuming that all of them
are vanishing one can show that (\ref{xabx}) is true.
 From eq. (\ref{xabx}) and (\ref{yay}) we conclude that the invariants ${\bf
x}_{ab}$ and ${\bf y}_{a}$ depend locally on $\bf x$, $\bf y$. More
precisely,
 \[ {\bf x}_{ab}=2(\rho^{-2}{\bf x})_{,ab}
  -\eta_{ab}(\rho^{-2}{\bf x})^{,c}{_c} \]
 \[{\bf y}^a=- 2E^{ab}(\rho^{-2}{\bf y})_{,b} \rho^{2} \]
and the primary data $({\bf x},{\bf y})$ fulfills usual wave equation.

\[ (\rho^{-1}{\bf y})^{,a}{_a}+\rho\dtwo {\bf y}=0 \quad
(\rho^{-1} {\bf x})^{,a}{_a}+\rho\dtwo{\bf x}=0 \]
We describe in the
sequel how the full Riemann tensor can be reconstructed from the
invariants~${\bf x},{\bf y}$.

\subsection{Quasi-local relations between gauge invariants
 and linearized Riemann or Weyl tensor}\label{QNP}

It is convenient to use skew-symmetric tensor $\varepsilon^{ab}$
instead of density $E^{ab}$. It can be defined as follows
\[ \sqrt{|\det\eta_{ab}|}\varepsilon^{uv}=1 \, ; \quad
{\varepsilon_{uv}\over \sqrt{|\det\eta_{ab}|}}=-1 \quad
\varepsilon^{ab}=2E^{ab} \]
 One can show that the linearized Riemann
tensor has the following $(2+2)$ ``spherical decomposition''. In terms of
our invariants it decouples into axial part

\[ \frac12 \rho^{-2}\varepsilon^{ab}\varepsilon^{CD}R_{abCD}={\bf y} \]

\[  R_{abcD||E}\varepsilon^{ab}\varepsilon^{DE} =\rho^3(\rho^{-1}{\bf y})_{,c} \]

\be\label{apR}
 \varepsilon^{AB}R_{AB}{^C}{_{d||C}}= \varepsilon_d{^b}
\rho^3(\rho^{-1}{\bf y})_{,b} \ee

\[ 4\rho^{-4}\kolo{R}^A{_{bCd||AD}}\varepsilon^{CD}
 =(\rho^{-2}{\bf y}_{d})_{,b} + (\rho^{-2}{\bf y}_{b})_{,d}  \]
and polar part

\[ \rho^{-2}\eta^{AC}\eta^{BD}R_{ABCD}=\frac12
\rho^{-2}\varepsilon^{AB}\varepsilon^{CD}R_{ABCD} ={\bf x} \]

\[ -2\rho^{-2} \varepsilon^{ab} \varepsilon^{cd}R_{abcd}= \rho^{-2}\eta^{ac}\eta^{bd}
R_{abcd} =- \rho^{-2}\eta^{ac}\eta^{BD} R_{aBcD} ={\bf x} \]

\[ \rho^{-2}\varepsilon^{AB}R_{ABCd||E}\varepsilon^{CE}={\bf x}_{,d} \]

\be 4\rho^{-4}\kolo{R}^A{_b}{^C}{_{d||AC}}=-{\bf x}_{bd} \ee

\[ \rho^{-2}(\dtwo +2)R_{abc}{^D}{_{||D}} \varepsilon^{ab}=-\rho^{-1}
(\rho{\bf x}_{ac})_{,b}\varepsilon^{ab} \]

\[ 2\rho^{-2}(\dtwo +2)R^A{_{bAd}} ={\bf x}_{bd}+
   (\rho^{-2}{\bf x}_{,d})_{,b} + (\rho^{-2}{\bf x}_{,b})_{,d}  \]
In above formulae, as in the whole paper, we use extensively some
operators on a unit sphere which become isomorphisms when we assume that
mono-dipole part of the field vanishes (see here Appendix \ref{multipole}
and also \cite{Beig}
or Appendix B in \cite{cjm}). The above equations contain the full
information on ten independent components of the Weyl tensor up to the
mono-dipole part of the field\footnote{ The mono-dipole part is
disscussed in \cite{JJspin2} and it corresponds to the charges related
to the Poincar\'e group.}. Moreover, one can easily check the
``peeling'' property \cite{Penrose} at $\scri^+$ starting from the
invariants ${\bf x},{\bf y}$ as a primary data. More precisely,
assuming the following expansion \be {\bf x} = {\bf x}_1 \rho + {\bf x}_2 \rho^2 + {\bf
x}_3 \rho^3 +{\bf x}_4 \rho^4+ \ldots \ee and the same form for $\bf y$
\be {\bf y} = {\bf y}_1 \rho + {\bf y}_2 \rho^2 + {\bf y}_3 \rho^3
+{\bf y}_4 \rho^4 + \ldots \ee
 we have
\[ {\bf x}_u = -\dot{\bf x}_1 \rho + (\frac12 {\bf x}_1
-\dot{\bf x}_2) \rho^2 + \ldots \]
\[ {\bf x}_v = \frac12 {\bf x}_1 \rho^2 - \frac12 {\bf x}_3 \rho^4 -
{\bf x}_4\rho^5  + \ldots \]
\[ {\bf x}_{uu} = 2\ddot{\bf x}_1 \rho^{-1} + 2(\ddot{\bf x}_2
 -\dot{\bf x}_1) +\ldots \]
\[ {\bf x}_{vv} = \frac12 {\bf x}_3 \rho^{3} +3{\bf x}_4\rho^4
  +\ldots \]
   We summarize below in the table
the relation of our invariants with the Newman-Penrose \cite{NP} scalars and the
Christodoulou-Klainerman-Nicol\`o \cite{CKN} decomposition of the Weyl
tensor:
\[ \begin{array}{|l|c|c|c|c|c|c|}
\hline {\rm Price} & {\rm Weyl} & {\rm C-K-N} &{\rm polar} &{\rm axial}
& {\rm N-P} & {\rm asymptotics} \\ \hline \Psi_2 & \kolo{W}_v{^A}{_{vB}}
& \rho^{-2}{\alpha}_{\scriptscriptstyle\rm CKN} & \rho^2{\bf x}_{vv} &
\rho^2(\rho^{-2}{\bf y}_{v})_{,v} & \Psi_0 & \rho^5 ({\bf x}_3 , {\bf y}_3) \\[0.5ex]
\Psi_{-2} & \kolo{W}_u{^A}{_{uB}}
&\rho^{-2}\underline{\alpha}_{\scriptscriptstyle\rm CKN} & \rho^2{\bf
x}_{uu} &
\rho^2(\rho^{-2}{\bf y}_{u})_{,u} & \Psi_4 & \rho (\ddot{\bf x}_1 , \ddot{\bf y}_1)\\[2ex]
\Psi_{1} & \rho^{-1} \varepsilon^{ab}{W}_{abv}{^A} &
\rho^{-1}{\beta}_{\scriptscriptstyle\rm CKN} &
 \rho^2(\rho^{-1}{\bf x})_{,v} &
\rho^2(\rho^{-1}{\bf y})_{,v} & \Psi_1 & \rho^4 ({\bf x}_2 , {\bf y}_2)\\[2ex]
\Psi_{-1} & \rho^{-1} \varepsilon^{ab}{W}_{abu}{^A} &
\rho^{-1}\underline{\beta}_{\scriptscriptstyle\rm CKN} &
\rho^2(\rho^{-1}{\bf x})_{,u} &
\rho^2(\rho^{-1}{\bf y})_{,u} & \Psi_3 & \rho^2 (\dot{\bf x}_1 , \dot{\bf y}_1)\\[2ex]
\Psi_{0} & W^a{_{bcd}}, W^A{_{Bcd}} &
\rho_{\scriptscriptstyle\rm CKN}, \sigma_{\scriptscriptstyle\rm CKN}  &
\rho^2{\bf x} & \rho^2 {\bf y} & \Psi_2 & \rho^3 ({\bf x}_1 , {\bf
y}_1)\\ \hline
\end{array} \]
where $\kolo{W}{^A}{_{abB}}={W}{^A}{_{abB}}-\frac12\delta^A{_B}
{W}{^D}{_{abD}}$ (cf. $TS$ transformation in appendix) and
${W}{^\nu}{_{\mu\lambda\delta}}$ is the linearized Weyl tensor.

One can easily check that among all $\Psi$'s only
$\rho^{-2}\Psi_0{^{\rm\scriptscriptstyle Price}}=
\rho^{-2}\Psi_2{^{\rm\scriptscriptstyle NP}}={\bf x}+i{\bf y}$
fullfills usual wave equation. The mono-dipole-free parts of the
invariants $\bf x$, $\bf y$ correspond to the unconstrained degrees of
freedom of the linearized gravitational field. Moreover, they play a
natural role of the ``positions'' in the reduced initial
data set on a Cauchy surface (\cite{GRG}, \cite{APP}, \cite{ISMC98}).\\
\underline{Remark} The Teukolsky equations \cite{Teukolsky} for
$\Psi_0,\Psi_4$ on a Kerr background seem to be quite strange as
primary equations because they are not deformations of the usual wave
equation when we pass to the asymptotically flat region. We would like
to stress that there exists a generalization for the notion of $\bf x$
and $\bf y$ on a Schwarzschild background and both invariants fulfill a
deformed wave equation -- Regge-Wheeler for $\bf y$ and Zerilli for
$\bf x$ (see \cite{ISMC98}).

\begin{Theorem}
The linearized Riemann tensor for the vacuum Einstein equations depends
quasilocally on the invariants $({\bf x},{\bf y})$ which contain the full
information about the linearized gravitational field. Moreover, the
invariants $\bf x$ and $\bf y$ fulfill usual wave equation.
\end{Theorem}

In other words any mono-dipole-free solution $({\bf x}, {\bf y})$ of
the wave equation gives a Weyl field:

\[ W_{abcd}=-\frac12\rho^2{\bf x} \varepsilon_{ab}\varepsilon_{cd} \]
\[ W_{ABcd}=-\frac12\rho^2{\bf y} \varepsilon_{AB}\varepsilon_{cd} \]
\[ W_a{^B}{_{cd||B}} =-\frac12 \varepsilon_{cd} \varepsilon^b{_a}
\rho^3(\rho^{-1}{\bf x})_{,b} \] \be \label{weyl}
W{_{aBcd||E}}\varepsilon^{BE} =-\frac12 \varepsilon_{cd}
\rho^3(\rho^{-1}{\bf y})_{,a} \ee
\[ \kolo{W}_c{^{AB}}{_{d||AB}}=\frac14 \rho^4 {\bf x}_{cd}=\frac12
\rho^4 \left[ (\rho^{-2}{\bf x})_{,cd} -\frac12\eta_{cd} (\rho^{-2}{\bf
x})^{,b}{_{b}}\right] \]
\[ \kolo{W}_c{^{A}}{_{Bd||AC}} \varepsilon^{BC}=\frac14
\rho^4 \left[ \varepsilon_c{^b}(\rho^{-2}{\bf y})_{,bd} +
\varepsilon_d{^b}(\rho^{-2}{\bf y})_{,bc}  \right]
\]
 and
${W}{^\nu}{_{\mu\lambda\delta}}$ fulfils field equations given by
Bianchi identities. The formulae (\ref{weyl}) are also valid (outside
origin) if we include mono-dipole part of the fields $\bf x$ and $\bf
y$ (see \cite{JJspin2}). More precisely, a mono-dipole solution ${\bf
x}=4{\bf m}\rho +12{\bf k}\rho^2$, ${\bf y}= 12{\bf s}\rho^2$
corresponds to the 10 Poincar\'e charges (a monopole and three
dipoles): $\bf m$ --- mass, $\bf s$ --- spin, $\bf k$ --- center of mass
and $\bf p$ --- linear momentum which is related with center of
mass by the relation ${\bf p} = (\partial_u +\partial_v){\bf k}$.
Moreover, the ``charges'' fulfill the following equations:
$\partial_a{\bf m}=\partial_a{\bf s}=\partial_a{\bf p}=0$, $\dtwo{\bf
m}=(\dtwo +2){\bf p}=(\dtwo +2){\bf
s}=(\dtwo +2){\bf k}=0$ which simply mean that $\bf m$ is a constant
and $\bf p$, $\bf s$ are constant dipoles.\\
\underline{Remark} One can check that the ``exterior'' bounds  (10.19
-10.20) for the various null components of the Weyl field in \cite{CKN}
(p. R114-R115) are related to the most important radiative terms ${\bf
x}_1$ and ${\bf y}_1$ which appear in
$\underline\alpha_{\scriptscriptstyle\rm CKN}$,
$\underline\beta_{\scriptscriptstyle\rm CKN}$,
$\rho_{\scriptscriptstyle\rm CKN}$ and $\sigma_{\scriptscriptstyle\rm
CKN}$ but the fall-off conditions on $\alpha_{\scriptscriptstyle\rm
CKN}$ and $\beta_{\scriptscriptstyle\rm CKN}$ are slightly weaker. In
particular it is not known how to improve the fall-off for
$\alpha_{\scriptscriptstyle\rm CKN}$ in the case of curved space-time.
This leads to the problem in the standard conformal approach where
stronger fall-off on $\alpha_{\scriptscriptstyle\rm
CKN}$\footnote{corresponding to $\Psi_0^{\rm NP}$} is assumed.

\subsection{Hierarchy of asymptotic solution on scri for scalar wave
equation}\label{hass}
Let us consider the wave equation in null coordinates $(u,v)$
 \be\label{wave}
  \rho^{-1}(\rho^{-1}\varphi)^{,a}{_a} +\dtwo \varphi=0 \ee
and suppose we are looking for a solution of the wave equation
(\ref{wave}) as a series (see \cite{FGF})
\be\label{szereg} \varphi = \varphi_1 \rho + \varphi_2 \rho^2 + \varphi_3
\rho^3 +\ldots \ee
where each $\varphi_n$ is a function on $\Scri^+$, $\partial_v \varphi_n =0$.

If we put the series (\ref{szereg}) into the wave equation (\ref{wave}),
we obtain the following recursion
\be\label{fir}
 \partial_u \varphi_{n+1}=-\frac1{2n}[\dtwo +(n-1)n]\varphi_n \, .\ee
The above formula is the same as equations 2, 3, 4 in \cite{BVM} and equation 1.6 in \cite{FGF}
but written in a more elegant way. Let us finish this Section with the following
Remark which is devoted to so-called {\em Newman-Penrose constants}: \\[1ex]
\indent \underline{Remark}. The kernel of the operator $[\dtwo  +
l(l+1)]$ corresponds to the $l$-th spherical harmonics. The right-hand
side of (\ref{fir}) vanishes on the $n-1$ spherical harmonics subspace.
This means that the corresponding multipole in $\varphi_{n+1}$ does not
depend on $u$. In particular, for $n=3$ we have quadrupole charge in
the fourth order. The nonlinear counterpart of this object is called
{\em Newman-Penrose constant} (see \cite{NP}, \cite{Burg}, \cite{APP}).
In particular, in our case, this constant is related to the quadrupole
part of ${\bf x}_4$ and ${\bf y}_4$.

\section{Gravitational field on a Schwarzschild background}
\label{lfS}
In \cite{ISMC98} it is shown how to generalize the gauge-invariant
quantities ${\bf x}$ and ${\bf y}$ for the case of Schwarzschild
background. We would like to investigate the following problems:\\
1. What is the relation (similar to (\ref{weyl}))
 between linearized Weyl tensor and the gauge
invariants?\\ and \\
2. How this relation applied to the
 asymptotics for the solutions of deformed wave equation
(Regge-Wheeler and Zerilli)
clarifies a peeling property for the linearized Weyl tensor?

In this section we summarize results of Adam Jankowski \cite{adamj}
who analyzed axial degree of freedom.
We shall often use the same letters for the generalizations of the
objects from flat to Schwarzschild background with the obvious
identification when the mass $m=0$.

\subsection{Schwarzschild metric in null coordinates}
 Let us start with spherical coordinates $ r,t, \theta ,
\phi$, hence the Schwarzschild metric has the following form:
\be \eta_{\mu\nu} dx^\mu dx^\nu =
-\left(1-\frac{2m}{r}\right)dt^2 + \frac{1}{1-\frac{2m}{r}} dr^2 + r^2
\left( d\theta^2 + \sin^2 \theta \, d \phi^2 \right)
\end{equation}
Spacetime  $M$ has a natural foliation with respect to spherical symmetry,
more precisely, it splits into
 $ S^2 \times M_2 $, where $ S^2 $ is a two-dimensional sphere
and $ M_2 $ is described by coordinates $r $ and $t$.
This way spacetime $ ( M , \eta ) $
with coordinates $ ( u,v,\theta, \phi) $ splits into
 $(S^2, \eta |_{S^2})$ and $(M_2 , \eta|_{M_2})$.
we shall often denote by $x^A$ coordinates $(\theta, \phi)$ on a
sphere,
but on $ M_2 $ we use null coordinates defined in terms of standard
 $ r $ and $t $ as follows: \ber
 u & = & t - r - 2m \ln(r-2m)  \hspace{.5in}  u \in ]
 - \infty , + \infty [
 \nonumber \\ \label{u,v}
 v & = & t + r + 2m \ln(r-2m)  \hspace{.5in}  v \in ] - \infty , + \infty [
\eer

Radial curves ( $\theta={\rm const.}$, $\phi={\rm const.}$) with fixed
 $u={\rm const.}$ are null geodesics, similarly for $v={\rm const.}$

\begin{center}
\includegraphics[width=14cm]{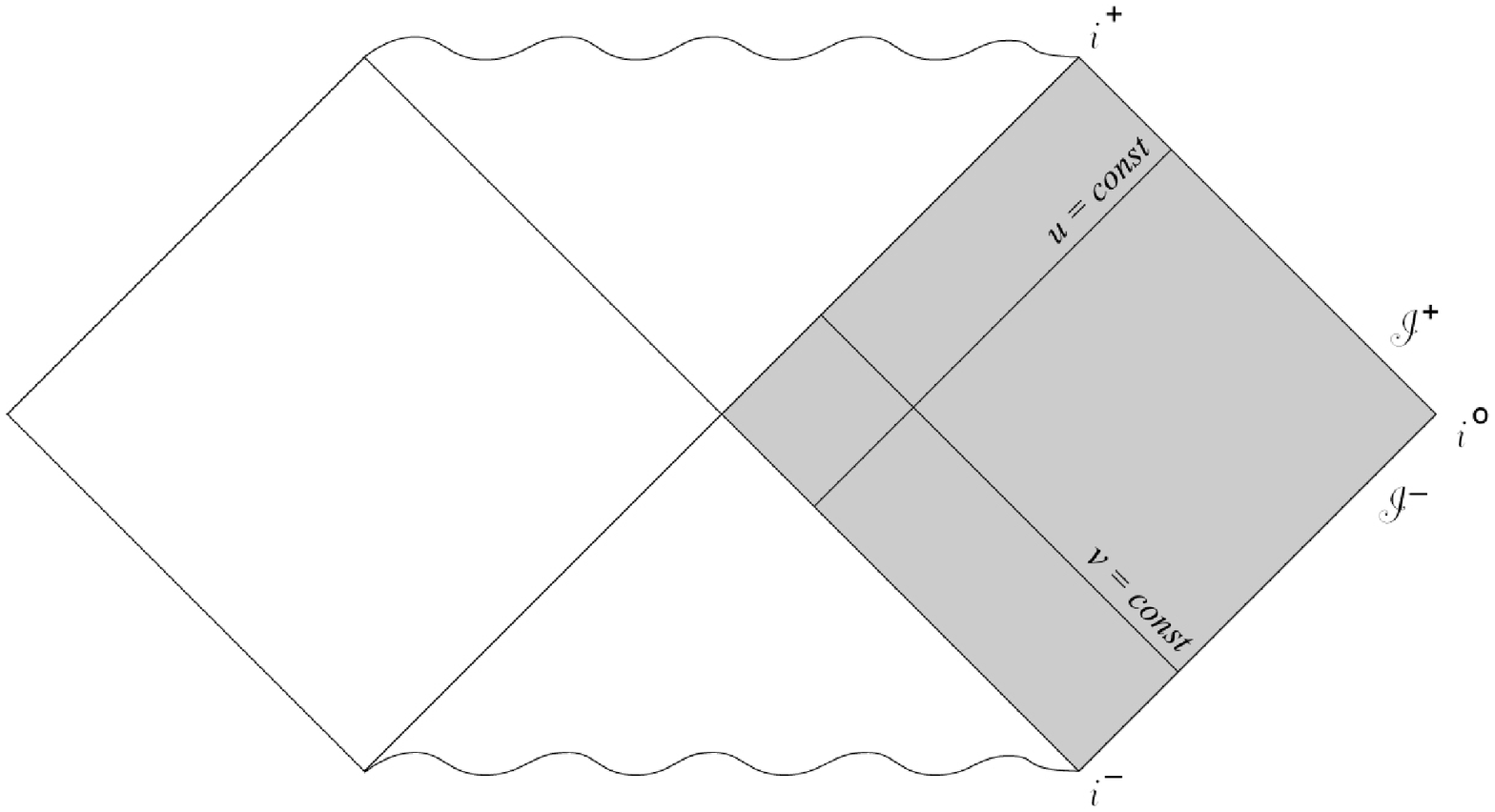}
\end{center}
Our main interest will concentrate on the domain $ r>2m $
marked by grey colour on the above {\it Carter--Penrose diagram for
Schwarzschild spacetime}.

Let us denote coordinates $(u,v)$ by small Latin characters $(x^a)$,
spherical coordinates by capitol Latin characters
 $ (x^A) $,$(A = 1, 2 )$, $(x^1 = \theta , x^2
= \phi )$, and the unit round spherical metric by
$\stackrel{\circ}{\gamma} _{AB}$:
\[
\stackrel{\circ}{\gamma} _{AB} dx^A dx^B = d\theta^2 + \sin^2\theta
d\phi^2 \, .
\]
Let us also notice that $M_2 \perp S^2$, because $ \eta_{aA} = 0$.

Let us fix some more notation:
symbol ``$||$'' denotes two-dimensional covariant
derivative on $S^2$ compatible with induced metric $\eta_{AB}$,
the equality \[
\eta_{AB} = r^2 \stackrel{\circ}{\gamma} _{AB}
\]
 implies that
the Christoffel symbols for $\eta_{AB}$ and $\stackrel{\circ}{\gamma} _{AB}$
are the same. Moreover, symbol
$\stackrel{\circ}{\triangle}$ denotes Beltrami-Laplace operator
for the unit metric $\stackrel{\circ}{\gamma} _{AB}$.
We keep the same notation as in Section \ref{ncfb} with obvious
generalization from flat background to Schwarzschild one.

 \ber \eta_{\mu\nu} dx^\mu dx^\nu & = & \eta_{ab}
dx^a dx^b + \eta_{AB} dx^A dx^B \eer  \ber
\label{metryka u,v} \eta_{\mu\nu} dx^\mu dx^\nu & = & -\left(
1-\frac{2m}{r}\right) du dv + r^2 ( d \theta^2+ \sin^2 \theta d\phi^2)
\eer
From definition (\ref{u,v}) we have $ \frac12(dv - du) =
k^{-1} dr $ where by $k$ we denote:
\[ k := 1-\frac{2m}{r} \, .\]
 We shall analyze the region far away from the sources and assume
 that $r > 2m$ which implies $ k > 0$.
 Using (\ref{u,v}) one can verify that
\begin{equation} \label{poch_r}
r_{,a} = - \ep_a \sqrt{k} \, ,
\end{equation}
where $\ep_a$ is defined as follows: \ber
\ep_v = -\pl k^\pl & &\ep_u = \pl k^\pl \\
\ep^v = - k^{-\pl} & &\ep^u = k^{-\pl}  \eer
and the indices are raised by the two-dimensional
inverse metric $\eta^{ab}$.
This way we have introduced a unit-length radial vector $\ep^a$
\[
\eta_{ab} \ep^a \ep^b = \eta^{ab} \ep_a \ep_b = 1 \, .
\]

One can also check that
\be \eta_{AB , a} = - \dwaskr \ep_a \eta_{AB} \ee
and we have the following nonvanishing Christoffel symbols for
the metric connection on  $M$:
 \be \label{gamma1} \Gamma^A{_{Ba}}= - \delta^A_B
\frac{\sqrt{k}}{r} \ep_a \, , \ee
\be \label{gamma2} \Gamma^a{_{AB}}=
\eta_{AB} \frac{\sqrt{k}}{r} \ep^a \, , \ee
\be \label{gamma3}
\Gamma^A{_{BC}}= \frac{1}{2} \eta^{AD}(\eta_{DB,C}+\eta_{DC,B}-
\eta_{CB,D}) \, , \ee
\be \label{gamma4} \g^a{_{bc}} = \frac{m}{r^2}
k^{\frac{1}{2}} ({-\delta^a}_b \varepsilon_c {-\delta^a}_c
\varepsilon_b +\varepsilon^a \eta_{bc}) \, . \ee
Let us notice that $\g^A{_{BC}}$ are simultaneously
 Christoffel symbols for
the induced two-dimensional metric $\left. \eta \right|_{S^2}$.
More precisely, in usual angular coordinates $(\theta, \phi)$
 we have two nonvanishing
components: $\Gamma^{\phi}{_{\phi\theta}}=\cot \theta$ and
$\Gamma^{\theta}{_{\phi\phi}} = -\sin\theta\cos\theta$. \\
 We can assign the symbols $\Gamma^A{_{BC}}$ and $\g^a{_{bc}}$  to
  covariant derivative on $S^2$
 and $M_2$ respectively, and denote (on the example of a covector)
  as follows
\be\label{Scov}
\xi_{a \tr b} := \partial_b \xi_a  - \xi_f {\g^f}_{ab}
\ee
\be\label{Mcov}
\xi_{A||B} := \partial_B \xi_A - \xi_F {\g^F}_{AB}
\ee
 Riemann tensor for the metric $\eta_{\mu\nu}$ equals to Weyl
 (Schwarzschild metric is a vacuum solution) and has
 the following nonvanishing components (up to the symmetries):
\begin{equation} \label{C1}
C^A{_{BCD}} = \frac {2m} {r^3} (\delta^A_C \eta_{BD} - \delta^A_D
\eta_{BC})\,,
\end{equation}
\begin{equation} \label{C2}
C^a{_{bcd}} = \frac {2m} {r^3} (\delta^a_c \eta_{bd} - \delta^a_d
\eta_{bc})\,,
\end{equation}
\begin{equation} \label{C3}
C^A{_{aBb}} = - \frac {m} {r^3} \delta^A_B \eta_{ab}\,.
\end{equation}
 Levi-Civita connection on $S^2$ has the following
 components of the two-dimensional Riemann tensor:
\begin{equation} \label{RnaS2}
 \stackrel{2}{R}{^A}{_{BCD}} =
 \frac1{r^2} (\delta^A{_C} \eta_{BD} - \delta^A{_D}
 \eta_{BC})\,.
\end{equation}
Similarly, the corresponding curvature for the metric  $\eta_{ab}$
has the form:
\begin{equation} \label{RM2}
\stackrel{2}{R}{^a}{_{bcd}} = -\frac{2m}{r^3} (\delta^a{_c}
\eta_{bd} - \delta^a{_d} \eta_{bc})\,.
\end{equation}
Let us introduce Levi-Civita tensor for $(S^2, \eta
|_{S^2})$ and $(M_2 , \eta|_{M_2})$ respectively:
\[ \ep^{AB}:=\frac{ E^{AB}}{\sqrt{ \left| \det \eta_{CD} \right| }}
\quad \ep^{ab}:=\frac{ E^{ab}}{\sqrt{ \left| \det \eta_{cd} \right| }}
\]
where $E^{ab}$ are coordinates of tensor density
$\frac{\partial}{\partial u} \wedge
\frac{\partial}{\partial v}$ so the values are in set
$\{-1,0,1\}$. Similarly,  $E^{AB}$ are coordinates for
$\frac{\partial}{\partial \theta} \wedge \frac{\partial}{\partial
\phi}$,
 so we may write explicitly
\be \label{t_levicivityAB}
 \ep^{AC}=\frac{E^{AB}}{r^2 \sin{\theta}} \,,
\ee
\be \label{t_levicivityab}
 \ep^{ab} = \frac{2 E^{ab}}{k} \,.
\ee
Let us notice that the metric $\eta_{ab}$ has a signature
$(+,-)$, hence its determinant is negative:
\[ \det \eta_{ab} < 0 \, .\]
This is important when we raise indices in
$\ep_{ab}$:
\be
 \label{ricci1} \ep_{ab} = \eta_{ac} \eta_{bd}
\ep^{cd}  = - \sqrt{ \left| \det \eta_{ab} \right| }
E_{ab} = - \pl k E_{ab}\,,
 \ee
 but on $S^2$ the corresponding sign is positive:
 \be
  \ep_{AB} = \eta_{AC}
\eta_{BD} \ep^{CD}  = \sqrt{ \left| \det \eta_{AB}
\right| } E_{AB} = r^2 \sin \theta E_{AB}\,.
 \ee

\noindent It would be useful 
to derive explicit formulae for first and second
derivatives of $\ep^{AC}$ and $\eta^{AC}$ with respect
to null coordinates on $M_2$ which are implied by (\ref{poch_r}):
\be \label{1 poch ep_AB}
{\ep^{AC}}_{\tr a} = \dwaskr \ep^{AC} \ep_a \, ,
\quad {\eta^{AC}}_{\tr a} = \dwaskr \eta^{AC} \ep_a
\ee
\be \label{2 poch ep_AB}
{\ep^{AC}}_{\tr ab} = (\frac{6k}{r^2} \ep_a \ep_b - \frac{2m}{r^3} \eta_{ab})
\ep^{AC} \, ,
\quad {\eta^{AC}}_{\tr ab} = (\frac{6k}{r^2} \ep_a \ep_b - \frac{2m}{r^3} \eta_{ab})
\eta^{AC} \, ,
\ee
where obviously the objects $\ep^{AC}$ and $\eta^{AC}$ are scalars
with respect to the covariant derivative (\ref{Mcov}).

\subsection{Gauge transformation for the linearized
metric tensor $h_{\mu \nu}$}
We shall analyze, from $2+2$ decomposition point of view,
the gauge transformation generated by infinitesimal diffeomorphism of
$M$ for the linearized metric tensor\footnote{The standard linearization
formulae have been shifted to the Appendix \ref{lRt}.}:
 \be \label{gauge} h_{\mu \nu} \rightarrow h_{\mu \nu} +
2\xi_{(\mu;\nu)} \, .\ee

Let us first split the covariant derivatives of the covector field
 $ \xi_\mu = ( \xi_a , \xi_A )$:
\[
\xi_{a;b} = \partial_b \xi_a - \xi_f \g^f{_{ab}} = \xi_{a \tr b}
\]
\[
\xi_{a;A} = \partial_A \xi_a - \xi_F \g^F{_{aA}} = \xi_{a||A} + \skr
\xi_A \ep_a
\]
\[
\xi_{A;b} = \partial_b \xi_A - \xi_F \g^F{_{bA}}= \xi_{A \tr b} + \skr
\xi_A \ep_b
\]
\[
\xi_{A;B} = \partial_B \xi_A - \xi_F \g^F{_{AB}} - \xi_f \g^f{_{AB}}  =
\xi_{A || B} - \skr \xi_f \ep^f
\]
and next apply them to gauge transformation of the tensor $h_{\mu\nu}$:
\be \label{cech 0}
h_{ab} \rightarrow h_{ab} + \xi_{a\tr b} +\xi_{b\tr a}\, ,
\ee
\be \label{cech 1} h_{aA} \rightarrow h_{aA} + \xi_{a||A} + \xi_{A\tr a}
+\dwaskr \ep_a \xi_A \, ,\ee
\be \label{cech 2} h_{AB} \rightarrow h_{AB} +
\xi_{A||B} + \xi_{B||A} - \dwaskr \xi_e \ep^e \eta_{AB} \, .\ee
Here the symbol
$\circ$ over two-dimensional tensor $t_{AB}$
denotes its traceless part:
\be \label{btr} \stackrel{\circ}{t}_{AB} := t_{AB} - \pl \eta_{AB}
\eta^{CD}t_{CD} \, .\ee
In particular, we denote the traceless part of
 $h_{AB}$  by \be \chi_{AB} \equiv
\stackrel{\circ}{h}_{AB} := h_{AB} - \pl \eta_{AB} H \, ,\ee
where $ H:= \eta^{AB}h_{AB} $.\\
From (\ref{cech 2}) we get the gauge transformation for
$\chi_{AB}$:
\[
\chi_{AB} \rightarrow \chi_{AB} + \xi_{A||B} + \xi_{B||A} - \eta_{AB}
{\xi^C}_{||C} \, .
\]
Let us notice that gauge of $\chi_{AB}$ depends only on $\xi_A$,
hence it is not dependent on the part which is tangent to $M_2$.

The ten components of the tensor $h_{\mu \nu}$ split naturally
with respect to the $2+2$--splitting of the spacetime $M=S^2 \times M_2$ into:
\begin{itemize}
\item components $ h_{ab} $ in $ M_2$,
\item components $ h_{AB} $ on $ S^2$,
\item mixed components  $ h_{aA}$.
\end{itemize}
However, for the description of the two degrees of freedom of
the gravitational field one can divide ten components of
the tensor $h_{\mu \nu}$ differently, into axial and polar part.
They split as follows:
\begin{itemize}
\item 7 polar components:  $h_{ab}$,  $h_a{^A{_{||A}}}$, $\chi^A {_{B||AC}} \eta^{BC}$,
$H$;
\item 3 axial components: $h_{aA||B} \ep^{AB}$, $\chi^A {_{B||AC}}
\ep^{BC}$.
\end{itemize}
In this section we shall consider only axial part of the gravitational
field. In particular, the gauge transformation for the axial components
of the tensor $h_{\mu \nu}$ reduces to:
 \be \label{cechowanie1}
h_{aB||C} \ep^{BC} \rightarrow h_{aB||C} \ep^{BC} + (\xi_{B||C}
\ep^{BC})_{\tr a} \, ,
 \ee
 \be \label{cechowanie2}
{{\chi_A}^B}_{||BC} \ep^{AC} \rightarrow {{\chi_A}^B}_{||BC}
\ep^{AC} + \frac{\lpd}{r^2}(\xi_{B||C} \ep^{BC})
  \ee
 (see also Appendix \ref{apendix_invariants}).
 Let us notice that the gauge of all axial components depends only on
  $\xi_A$ so they are not dependent on the infinitesimal change of
  coordinates on $M_2$. Moreover, we shall see in the sequel that
  using appropriate operators on $S^2$ one can produce from
  $h_{aB||C} \ep^{BC}$ and ${{\chi_A}^B}_{||BC}
\ep^{AC}$ a gauge invariant quantity.

\subsection{Gauge invariants}
The following object
 \ber \label{yaS}
 {\mathbf y}_a & := & \lpd( h_{aA||C} \ep^{AC}) - (r^2
{{\chi_A}^B}_{||BC} \ep ^{AC} )_{\tr a} \, .
 \eer
 is invariant with respect to the gauge transformation described
 in the previous subsection.
 It is useful to introduce another object
 (cf. \cite{ISMC98}):
 \ber \label{yS}
 {\bf y}   & := & r^2 (h_{bA||C} \ep^{AC}) _{\tr a} \ep^{ab} \, .
\eer
which is also gauge independent.
Let us notice that definitions (\ref{yaS}) and (\ref{yS}) are the
same as formulae (\ref{ya}) and (\ref{doy}) respectively. This
means that for axial and invariants there are no
``background mass'' corrections. This phenomena is no longer valid
for polar invariant $\bf x$ (see \cite{ISMC98}).
The explicit proof of gauge invariance property for
${\mathbf y}_a$ and ${\mathbf y}$
is given in appendix
\ref{apendix_invariants}.
The invariants ${\mathbf y}_a$ and ${\mathbf y}$
(introduced also in \cite{ISMC98}) are not
independent but they fulfill the following identity
 \be \label{zwiazek}
r^2 {\bf y}_{a \tr b } \ep^{ba}  =  \lpd {\bf y} \ee
which is a straightforward consequence of the above definitions of the objects.
The dipole part of ${\mathbf y}$ corresponds to the stationary solution
of the field equations\footnote{We shall introduce them
in the next subsection.}
(see \cite{ISMC98}). We shall verify this property in the sequel
analyzing asymptotics of the solutions at future null infinity
$\scri^+$.

\subsection{Field equations}
Let us consider linearized vacuum Einstein equations
\be \label{lricci}
 r_{\nu\rho} :=\frac12 \left( {h^{\sigma}}_{\rho;\nu\sigma}+{h^{\sigma}}_{\nu;\rho\sigma}-
{{h_{\nu\rho}}^{;\sigma}}_{\sigma}-{h^{\sigma}}_{\sigma;\nu\rho} +
h^\sigma{_\alpha}C^{\alpha}{_{\nu\sigma\rho}} \right) =0
\, .
 \ee

The axial part of linearized Einstein equations on a Schwarzschild
background can be
easily described in terms of invariants  \cite{adamj}
(we present the details in appendix \ref{apendix_rownania})
\be
\label{ryd} \ep_d{^a} ( r^2 {\mathbf y} )_{\tr a} + r^2 {\mathbf
y}_d = -2r^4 r_{Bd||D} \ep^{BD} = 0 \, ,
 \ee
\be \label{r2}
(r^2 {\mathbf y}^a)_{\tr a} = 2r^4 { \stackrel{\circ}{r} }_A{^B}_{||BC}
\varepsilon^{AC} = 0 \, .
\ee
 The above equations take
the same form as in the case of the flat background\footnote{This phenomenon is
valid {\em only} for axial part.}.
Moreover, they imply a second order hyperbolic equation for ${\mathbf y}$:
\[
(r^{-2}(r^2 {\mathbf y} )^{\tr a})_{\tr a} + r^{-2} \lpd {\mathbf y}  = 0 \, .
\]
However, this equation
is no longer a wave operator but Regge-Wheeler equation
\be \label{rRW}
( \Box + \frac{8m}{r^3}){\mathbf y}=0 \, .
\ee

\section{Gauge transformation of
linearized Riemann tensor and axial invariants}\label{stR}

In \cite{adamj} one can find details of calculations
which we would like to present in this section.
Those calculations are related to the gauge transformations of
 $r^\mu{_{\nu\rho\sigma}}:= \delta
R^{\mu}{_{\nu\rho\sigma}}$, $\delta
R_{\mu\nu\rho\sigma}$ and $ r_{\mu\nu}:= \delta R_{\mu\nu}$
(see appendix \ref{lRt}).
The linearized Ricci $r_{\mu\nu}$ is obviously gauge invariant
but $ r_{\mu\nu\rho\sigma} $ and $\delta
R_{\mu\nu\rho\sigma}$ are in general gauge dependent.

Using the formula (\ref{gauge}) one can show that
 the gauge transformation of the
linearized Riemann tensor (\ref{dR2}) discussed in appendix \ref{lRt}
takes the following form:

 \ber \label{gauge_R}
  \delta R_{\mu\nu\rho\sigma} & \rightarrow & \delta R_{\mu\nu\rho\sigma} +
           \xi_{\alpha}( {R^\alpha}_{\mu\sigma\rho;\nu} +
           {R^\alpha}_{\sigma\mu\nu;\rho} +
           {R^\alpha}_{\nu\rho\sigma;\mu} +
           {R^\alpha}_{\rho\nu\mu;\sigma}) + \nonumber \\ & &
           2 \xi_{\alpha;\mu} {R^\alpha}_{\nu\rho\sigma} +
           2 \xi_{\alpha;\rho} {R^\alpha}_{\sigma\mu\nu} +
           2 \xi_{\alpha;\sigma} {R^\alpha}_{\rho\nu\mu} +
           2 \xi_{\alpha;\nu} {R^\alpha}_{\mu\sigma\rho} \, .
\eer
 Let us denote the gauge term as
\[
 \delta R_{\mu\nu\rho\sigma} \rightarrow \delta
R_{\mu\nu\rho\sigma} + \gauge (\delta R_{\mu\nu\rho\sigma})
\]
 hence the corresponding components take the form:
 \ber
  \gauge(\delta R_{abED}) & = & 0 \, ,
 \eer
 \be
  \gauge(\delta R_{abeD})  =   \frac{2m} {r^3} (\xi_b \eta_{ae}
                     - \xi_a \eta_{be})_{||D} +  \frac{m}{r} \left(
                     \eta_{be} (\frac{\xi_D} {r^2})_{,a} -
                     \eta_{ae} (\frac{\xi_D} {r^2})_{,b}\right) \, ,
 \ee
 \ber
  \gauge(\delta R_{ABDd}) & = & \frac{m}{r^3}( \xi_{d||A} \eta_{BD} - \xi_{d||B} \eta_{AD}) +
  \frac{2m} {r^3} (\xi_B \eta_{DA} - \xi_A \eta_{DB} )_{\tr d} \nonumber \\
                & &  +\frac{4m \sk}{r^4} \ep_d( \xi_B \eta_{DA} -\xi_A
                \eta_{DB}) \, ,
 \eer
 \ber
  \gauge(\delta R_{AbCd}) & = & - \frac{m}{r^3}
  (\eta_{bd} \xi_{A||C} + \eta_{db} \xi_{C||A}) + \eta_{AC}
  F(\xi)_{db} \, ,
 \eer
 where
\[
F(\xi)_{db} :=   \frac{m}{r^2} \left[ (\frac{\xi_d}{r})_{\tr b} +
(\frac{\xi_b}{r})_{\tr d} + \frac{\sk}{2r^2}(\xi_d \ep_b +\xi_b
\ep_d ) - 2 \eta_{bd} \skr \ep^a \xi_a \right] \, .
\]
To construct axial part of the Riemann
 tensor $\delta R_{\mu\nu\rho\sigma}$ we use some
  ``spherical operators'' and obtain the following gauge dependence:

 \ber \label{zlram1}
  r^2 \delta R_{abeD||A}\ep^{DA} & \rightarrow &
  r^2 \delta R_{abeD||A}\ep^{DA}
   +\frac{m}{r^3} \left(\eta_{be}(\xi_{D||A}\ep^{DA})_{,a}
  - \eta_{ae}(\xi_{D||A}\ep^{DA})_{,b} \right) \, ,
 \eer
 \ber \label{zlram2}
  \delta R_{AB}{^C}{_{d||C}}\ep^{AB} & \rightarrow & \delta R_{AB}{^C}{_{d||C}}\ep^{AB} -
                     \frac{4m}{r^3} (\xi_{A||B} \ep^{AB})_{\tr d}
                     \, ,
 \eer
 \ber \label{zlram3}
  \stackrel{\circ}{\delta R}_{AbBd} & \rightarrow &
  \stackrel{\circ}{\delta R}_{AbBd} -
                     \frac{m}{r^3}( \xi_{A||B} + \xi_{B||A} - \eta_{AB}
                     \xi^D{_{||D}})\eta_{bd} \, ,
 \eer
 \ber \label{zlram4}
  \delta R_{abED} & \rightarrow & \delta R_{abED} \, .
 \eer
 Let us notice that the components of the linearized Riemann tensor
  (\ref{zlram1}--\ref{zlram3}) may be ``corrected'' in such a way that
  we obtain gauge-independent objects:
\[
r^2 \lpd \delta R_{abcD||A}\ep^{DA} \ep^{ab}+ \frac {4m}{r}
\ep^b{_{c}} (r^2 {{\chi_A}^B}_{||BC} \ep^{AC})_{\tr b} =
\]
\[
r^2 \lpd r_{abcD||A}\ep^{DA} \ep^{ab}+ \frac {4m}{r} \ep^b{_{c}}
(r^2 {{\chi_A}^B}_{||BC} \ep^{AC})_{\tr b} + \frac{m}{r^3}
h_{Da||A} \eta_{bc} \ep^{DA}
\]
\[
\delta R_{AB}{^C}{_{d||C}}\ep^{AB} + \frac{4m}{r^3} h_{dB||C}
\ep^{BC}  = r_{AB}{^C}{_{d||C}}\ep^{AB} + \frac{5m}{r^3} h_{dB||C}
\ep^{BC}
\]
\[
\stackrel{\circ}{\delta R}_{AbBd} + \frac{m}{r^3} \eta_{bd} \chi_{AB} =
\, \stackrel{\circ}{r}_{AbBd} + \frac{2m}{r^3} \eta_{bd} \chi_{AB}
\]
\[
\delta R_{abED} = r_{abED}
\]
For completeness, we give here also the above above formulae in terms of
the ``true'' linearized Riemann tensor $r_{\mu \nu \rho \sigma}$
which is related to $\delta\! R_{\mu \nu \rho \sigma}$ by formula
 (\ref{zwiazek_r_R}).
  More precisely, the relation (\ref{zwiazek_r_R})
  for various components of Riemann tensor gives the following:
\[
r_{abED}  =  \delta R_{abED} \, , \quad r_{AbED}  =  \delta R_{AbED}
\, , \quad
r_{ABEd}  =  \delta R_{ABEd} -  \frac {m} {r^3} h_{Ad} \eta_{BE}
\]
\[
r_{abeD}  =  \delta R_{abeD} -  \frac {m}{r^3} h_{Da} \eta_{be}
\, , \quad
r_{AbBd}  =  \delta R_{AbBd} - \frac{m}{r^3} h_{AB} \eta_{bd} \, .
\]
One can easily verify that the attempt to exchange
 $\delta
R_{\mu\nu\lambda\rho}$ with $r_{\mu\nu\lambda\rho}$
does not improve the invariance of (\ref{zlram1} --
\ref{zlram3}). In particular, the ``metric corrections''
although in a different form will still occur.
We prefer to use the tensor $\delta
R_{\mu\nu\lambda\rho}$ because it possesses all the
  symmetries of the usual full curvature tensor.

The explicit formulae for the axial part of
 $\delta R_{\mu \nu \rho\sigma}$ in terms of the metric $h_{\mu \nu}$
 \ber
 2\delta R_{abED} & = & h_{aD \tr b || E} + h_{bE \tr a || D}
- h_{aE \tr b || D}
- h_{bD \tr a || E} +  \\
         &  & + \skr \ep_b ( 2 h_{aD || E} -2 h_{a E || D} )
         + \skr \ep_a ( 2 h_{bE || D} -2 h_{b D || e} ) \nonumber
 \eer
 \ber
2\delta R_{ABEd} & = & h_{Ad||BE} - h_{Bd||AE} + h_{BE||A \tr d} - h_{AE||B \tr d} +\nonumber \\
&  & +\skr ( 2 h_{BE||A} \ep_d - 2 h_{AE||B} \ep_d + h_{ad||A} \ep^a \eta_{BE} - h_{ad||B}
\ep^a \eta_{AE}) +\nonumber  \\
&  & +\skr( h_{Bd \tr a} \ep^a \eta_{AE} - h_{Ad \tr a} \ep^a \eta_{BE}
+ h_{Aa \tr d} \eta_{EB} \ep^a - h_{Ba \tr d} \eta_{EA} \ep^a ) + \nonumber \\
&  & + 2 \frac {k}{r^2} \ep^a \ep_d ( h_{Aa} \eta_{EB} -  h_{Ba} \eta_{EA} )
 \eer
 \ber
2\delta R_{abeD} & = &  h_{aD \tr be} - h_{ae \tr b||D} - h_{bD\tr ae} + h_{be \tr a || D} +\\
&  & +\skr ( h_{aD \tr e} \ep_b - h_{bD \tr e} \ep_a + h_{De \tr
a} \ep_b - h_{De \tr b} \ep_a + h_{be||D} \ep_a - h_{ae||D} \ep_b)
\nonumber
 \eer
 \ber
 2\delta R_{AbBd} & = & h_{Ad \tr b || B} -
h_{bd ||AB} - h_{AB \tr bd} + h_{bB||A \tr d} +
\nonumber \\
 &  & +\skr ( h_{Ad||B } \ep_b + h_{bB||A } \ep_d - h_{bA||B } \ep_d -h_{Bd||A } \ep_b )+
 \nonumber \\
  &  & -\skr ( h_{AB \tr b} \ep_d + h_{AB \tr d} \ep_b + h_{fd \tr b } \ep^f \eta_{AB}
  + h_{bf \tr d } \ep^f \eta_{AB} + \nonumber \\
  &  & + h_{bd \tr f } \ep^f \eta_{AB})
    +\frac {m}{r^3} (h_{AB} \ep_b \ep_d - h_{AB} \eta_{db} - 2 h_{db} \eta_{AB} )
\eer
enables one to express them in terms of the invariants
$ {\mathbf y}_a$ and ${\mathbf y}$:
 \be \label{RmmDD}
 \frac12 r^2 \ep^{ab}
\ep^{ED} \delta R_{abED}  =  {\mathbf y} \ee
 \[
r^2 (\stackrel{\circ}{\triangle} +2) {{\delta R_{AB}}^C}_{d||C}
\ep^{AB} + \frac{4m}{r^3} (r^2 {{\chi_A}^B}_{||BC}\ep^{AC})_{\tr
d} = \frac1r\ep_d{^a} (\dtwo +2)(r{\mathbf y})_{\tr a}
- \frac{4m}{r^5} \ep_d{^a} (r^2{\mathbf y})_{\tr a} \]
 \be \label{RDDDm}
 \hspace*{6cm} =
(\stackrel{\circ}{\triangle} +2) {\mathbf y}_d + r \sk \ep^a (
{\mathbf y}_{a \tr d} - {\mathbf y}_{d \tr a})
- \frac{4m}{r^3} {\mathbf y}_d
 \ee
 \be \label{zle}
4 r^4 \left[ \stackrel{\circ}{\delta R}{^A}{_{bBd||AD}} \ep^{BD} +
\frac{m}{r^3} \eta_{bd}{{\chi_A}^B}_{||BC}\ep^{AC} \right]  =
 (r^2 {\mathbf y}_d)_{\tr b} + (r^2 {\mathbf y}_b)_{\tr d}
\ee
\[
r^2 \lpd \delta R_{abeD||A}\ep^{DA}\ep^{ab}
 + \frac {4m}{r} \ep_{e}{^b} (r^2 {{\chi_A}^B}_{||BC} \ep^{AC})_{\tr
b} = \]
 \be \label{RmmmD}
 \hspace*{8cm} = \frac{1}{r} [ r \lpd {\mathbf y}]_{\tr e}
 + \frac{2m}{r^3} (r^2 {\mathbf y} )_{\tr e} \, .
 \ee
The above formulae describe the relation
between axial part of the linearized
Riemann tensor (corrected to gauge-independent form)
and our standard gauge-independent quantities
${\mathbf y}$, ${\mathbf y}_{a}$. The equations
(\ref{RmmDD}--\ref{RmmmD}) are generalizations of the formulae
(\ref{apR}).

\section{Asymptotics for solutions of Regge-Wheeler equation
at null infinity}
Let us consider Regge-Wheeler equation (\ref{rRW}) in null
coordinates $(u,v)$
 \be \label{wave1}
- \frac{2}{k} \left[  \partial_u ( r^2 \partial_v {\mathbf y}) +
\partial_v ( r^2 \partial_u {\mathbf y}) \right] + \dtwo
{\mathbf y} + \frac{8m}{r^3} {\mathbf y}=0  \ee
and let us assume
(cf. \cite{FGF}) that the solution of (\ref{wave1}) is in the following
asymptotic form (see appendix \ref{RWs}):
 \be\label{szereg1}
  {\mathbf y} = \sum_{n=1}^5 \frac{a_n(u,\theta,\phi)}{r^n}
  + O \left(\frac{1}{r^6} \right)
 \ee
where $a_n$ are functions of ($u$, $\theta$, $\phi$) hence they are well
defined on $\scri^+$.
 From the assumption that $a_n$ do not depend on $v$ we have for
 each $n$:
\[ \partial_v a_n =0 \, .\]
Moreover, denote the $u$-derivative  by dot e.g.
 \[  \dot{a}_n = \partial_u a_n \, . \]
 Additionally, using (\ref{poch_r}) together with the definition
 of $\ep_a $ one can express the derivatives of $k$
and $r$
\[
\partial_v k = \frac{m}{r^2} k  \; , \quad
\partial_u k = -  \frac{m}{r^2} k \; , \quad  \partial_v r = \pl k \; ,
\quad \partial_u r = - \pl k \, . \]


The series (\ref{szereg1}) inserted into equation
(\ref{wave1}) gives the following formula:\footnote{We assume that
the derivatives of the asymptotic terms ${\rm O}\left(\frac{1}{r^6} \right)$
are at least of the same asymptotic order.}
 \be\label{fir1} 
 \sum_{n=1}^5 \left[ n(n-1) \frac {a_n}{r^n}
 + \frac {2(4-n^2)a_n}{r^{n+1}} m
 + 2(n-1) \frac{\dot{a}_n}{r^{n-1}} + \frac{\dtwo a_n}{r^n}
 \right]  + {\rm O}\left(\frac{1}{r^6} \right) = 0 \, .
 \ee
 Comparing the coefficients at the same power of $r$
 we obtain recurrence relations for the coefficients~$a_n$:
\ber
  n=1 & & a_1 \mbox{--- free data} \nonumber \\
  n=2 & & 2 \dot a_2 + \dtwo a_1 =0 \label{rek} \\ \nonumber
 n \geq 2 & &
2n \dot{a}_{n+1} + \left[ \dtwo + n(n-1) \right] a_n -2 m
a_{n-1}(n-3)(n+1) = 0
 \eer
Let us rewrite equation (\ref{rek}) using new integer
parameter $l := (n-1)$:
 \be \label{fir2}
 2(l+1) \dot{a}_{l+2} = - \left[ \dtwo + l(l+1) \right] a_{l+1} + 2 m
a_{l}(l-2)(l+2)
 \ee

\indent \underline{Remark}. The Remark from Subsection \ref{hass}
about NP constants remains valid for the Schwarz\-schild background.
More precisely, the right-hand
side of (\ref{fir2}) vanishes on the spherical harmonics subspace
corresponding to $l=2$ and quadrupole part of
 $a_4$ does not depend on $u$.

\subsection{``Peeling'' for the axial part of Weyl tensor}
We continue our asymptotic
considerations based on the assumption
  (\ref{szereg1}), in particular the first term of the asymptotics
  gives
\[ {\mathbf y} ={\rm O} \left( \frac{1}{r}\right) \, . \]
The equation (\ref{ryd})
written in an equivalent form:
\begin{equation} \label{asympt}
{\mathbf y}_d = -\frac{1}{r^2} \ep_d{^a} ( r^2 {\mathbf y})_{\tr
a}
\end{equation}
may be used to obtain the asymptotic behaviour of
${\mathbf y_a}$:
\begin{eqnarray}
(r^2 {\mathbf y}) _{\tr u} & = & r \dot{a}_1+ (\dot{a}_2 - \pl
a_1) + \frac{1}{r} (\dot{a}_3 +m a_1 ) + \frac{1}{r^2} (\dot{a}_4
+ \pl a_3 ) + \nonumber \\ & &  \frac{1}{r^3}(\dot{a}_5 - m a_3
+a_4) + {\rm O} \left(\frac{1}{r^4} \right) \label{asymp-yu}
\end{eqnarray}
 \be \label{asymp-yv}
(r^2 {\mathbf y}) _{\tr v}  = - \pl a_1 + \frac{1}{r} m a_1  +
\frac{1}{2 r^2} a_3 + \frac{1}{r^3}(- m a_3 +a_4) + {\rm O}
\left(\frac{1}{r^4} \right) \ee
or more explicitly, using formula
 (\ref{asympt}) we obtain asymptotics of both components of
 ${\mathbf y}_a$:
 \be
 {\mathbf y}_u = -\frac{1}{r^2} \ep_u{^u} ( r^2 {\mathbf y})_{\tr
u}  \sim -\frac{\dot a_1}{r}
 \ee
 \be
 {\mathbf y}_v = -\frac{1}{r^2} \ep_v{^v} ( r^2 {\mathbf y})_{\tr
v}  \sim -\frac{a_1}{2r^2} \, .
 \ee
From (\ref{RmmmD}) one can show
\be \label{blast} \hspace*{-0.5cm} r(\dtwo +2)\delta
R_{abcD||A}\varepsilon^{DA}\varepsilon^{ab} + \frac {4m}{r^2}
\varepsilon_c{^b} (r^2 {{\chi_A}^B}_{||BC} \varepsilon^{AC})_{, b}
 = \frac{1}{r^2} (\dtwo +2)(r {\bf  y})_{, c} + \frac{2m}{r^4} (r^2 {\bf
 y})_{,c}\, .
 \ee
 From asymptotics (\ref{szereg1}) for the solutions
 of the equation (\ref{rRW}) we conclude that the
 $v$-component of equation (\ref{blast}) has the following asymptotic
 behaviour
\[ \frac{1}{r^2} (\dtwo +2)(r {\bf  y})_{, v} + \frac{2m}{r^4} (r^2 {\bf
 y})_{,v} = {\rm O} \left( \frac1{r^4}\right) \, ,
\]
because $\displaystyle (r{\bf y})_{,v} \sim \left( \frac{a_2}r\right)=
 {\rm O} \left( \frac1{r^2}\right)$
and by the use of (\ref{asymp-yv}). Moreover, the asymptotic condition
$\displaystyle \partial_v \chi^A{_B} = {\rm O}(\frac1{r^2})$,
which is usually fulfilled for the asymptotically flat metric,
implies
\[
\frac {4m}{r^2} \varepsilon_v{^b} (r^2 {{\chi_A}^B}_{||BC}
\varepsilon^{AC})_{, b} = {\rm O} \left( \frac1{r^4}\right) \, .
\]
 Hence
 \be \delta\Psi_1 {^{\rm\scriptscriptstyle NP}} \rightarrow
  r \delta R_{abvD||A}\varepsilon^{DA}\varepsilon^{ab}
 = {\rm O} \left( \frac1{r^4}\right)
 \ee
  gives a usual ``peeling'' because the gauge-dependent term
 \[ \frac {4m}{r^2} \varepsilon_v{^v} (r^2 {{\chi_A}^B}_{||BC}
\varepsilon^{AC})_{, v} = {\rm O}\left(\frac1{r^4}\right)
 \] has the same asymptotics as the full invariant.

Similarly we may investigate the invariant
 \be \label{last}
 \left[
{\stackrel{\circ}{\delta R}}{^A}_{bBd} + \frac{m}{r^3} \eta_{bd}
\chi^A{_B}\right]_{||AC}\ep^{BC} r^2
 \ee
 which equals
 \be\label{rhs} \frac{1}{r^2} \left[ (r^2
{\mathbf y}_d)_{\tr b} + (r^2 {\mathbf y}_b)_{\tr d} \right] \,
 \ee
 from (\ref{zle}).
 Using (\ref{szereg1}), (\ref{asymp-yu}--\ref{asymp-yv})
we obtain asymptotics O$(\frac1{r^5})$ for the term (\ref{rhs})
if we assume that both indices $b=d=v$. Hence
 \be \delta\Psi_0{^{\rm\scriptscriptstyle NP}} \rightarrow
  r^2 \kolo{\delta R}_v{^A}{_{vB||AC}} \ep^{BC} = {\rm O}
 \left( \frac{1}{r^5} \right)
 \ee
 Here the result is simpler because
 $\kolo{\delta R}_{vAvB}$
 is already gauge-independent ($\eta_{vv}=0$).

This way we have proved  ,,peeling''
for axial part of the Weyl tensor
corresponding to  Newman-Penrose scalars $\Psi_0$ and $\Psi_1$.
In a similar way one can check the asymptotics for the
remaining three scalars
$\Psi_3$, $\Psi_4$ and $\Psi_2$. This is a consequence of the same
tensor
equations (\ref{blast}, \ref{last}) but we need to use other
components of them. The corresponding invariants have the following asymptotic
behaviour:
\[
 \frac{1}{r^2} (\dtwo +2)(r {\bf  y})_{, u} + \frac{2m}{r^4} (r^2 {\bf
 y})_{,u} = {\rm O} \left( \frac1{r^2}\right) \, ,
\quad
\frac{2}{r^2} (r^2 {\mathbf y}_u)_{\tr u}
 = {\rm O} \left( \frac1{r}\right) \, .
\]
It is relatively easy to verify that the gauge-dependent ``correction'' terms
depending on
 $\chi^A{_B}$ have the same asymptotic order and finally we obtain
 \be\label{psi3}
    \delta\Psi_3 {^{\rm\scriptscriptstyle NP}}  \rightarrow r \delta
R_{abuD||A}\varepsilon^{DA}\varepsilon^{ab}
  = {\rm O} \left(\frac{1}{r^2}\right) \, ,
 \ee
 \be
 \delta\Psi_4{^{\rm\scriptscriptstyle NP}}
\rightarrow \kolo{\delta R}_u{^A}{_{uB||AC}} \ep^{BC} r^2
  = {\rm O} \left(\frac{1}{r}\right) \, ,
 \ee
and for the last gauge-independent component
 \be \label{endofpeel}
 \delta\Psi_2{^{\rm\scriptscriptstyle NP}}
 \rightarrow
 \ep^{ab} \ep^{ED} \delta R_{abED}
  = {\rm O} \left(\frac{1}{r^3}\right) \, .
 \ee
We would like to stress that
asymptotic behaviour of $\Psi_2, \Psi_3,
\Psi_4$ given by (\ref{psi3}--\ref{endofpeel}) is not
ambiguous\footnote{It is shown in \cite{Kroon},
even for much weaker asymptotic assumptions ---
so called polyhomogeneous asymptotics (i.e. including terms
$W(\ln r)r^{-k}$ where $W$--polynomial),
that the same asymptotics is valid for $\Psi_2, \Psi_3, \Psi_4$
but for $\Psi_0$ and $\Psi_1$, in general, we have
terms of order $r^{-4}{\ln r}$.}.
However, in general,  $\Psi_0$ and $\Psi_1$ may have weaker asymptotic
behaviour \cite{CKN}.
It is clear from the above investigations that
axial part obeys ``strong peeling'', but this is completely
not obvious that the same is true for the second degree of freedom
described by the polar part.

The results of this section we summarize in the following table:
$$ \begin{array}{|c|c|c|c|l|}
\hline   {\rm Weyl} & {\rm C-K-N}
&\mbox{axial} \ \mbox{invariant} & {\rm N-P}
& {\rm asymptotics} \\ \hline   \kolo{W}_v{^A}{_{vB}} &
r^{2}{\alpha}_{\scriptscriptstyle\rm CKN} & {r^{-2}}  (r^2
{\mathbf y}_v)_{\tr v}  & \Psi_0 & r^{-5} \ (r^{-3.5} ) \\[0.5ex]
  \kolo{W}_u{^A}{_{uB}}
&r^{2}\underline{\alpha}_{\scriptscriptstyle\rm CKN} & {r^{-2}}
 (r^2 {\mathbf y}_u)_{\tr u} & \Psi_4 & r^{-1}  \\[2ex]
  r \varepsilon^{ab}{W}_{abv}{^A} &
r {\beta}_{\scriptscriptstyle\rm CKN} & {r^{-2}} (\dtwo +2)(r
{\bf y})_{, v} + \frac{2m}{r^4} (r^2 {\bf
 y})_{,v}& \Psi_1 & r^{-4} \ ( r^{-3.5}) \\[2ex]
  r \varepsilon^{ab}{W}_{abu}{^A} &
r \underline{\beta}_{\scriptscriptstyle\rm CKN} & {r^{-2}}
(\dtwo +2)(r {\bf y})_{,u} + \frac{2m}{r^4} (r^2 {\bf
 y})_{,u}& \Psi_3 & r^{-2} \\[2ex]
  \ep^{ab} \ep^{ED} W_{abED}&
 \sigma_{\scriptscriptstyle\rm CKN}  &
  r^{-2} {\bf y} & \Psi_2 & r^{-3} \\ \hline
\end{array} $$
Here $W$ corresponds to $\delta R$, and in brackets we give
the asymptotic results of Christodolou-Klainerman-Nicol\'o
(cf. \cite{CKN}).

\section{Conclusions}
In \cite{Price} one can find the following statement:
 ``For even waves, it has not yet been possible to derive an
equation like Regge-Wheeler from the perturbed NP
equations''. This question has been resolved in
\cite{F-L} but in our opinion not in a satisfactory way
(see the discussion at the end of this Section).
 We shall explain in a separate paper why in \cite{Price}
 one can  easily formulate Regge-Wheeler
equation for odd degree of freedom in terms of Newman-Penrose
scalars but for polar (even) degree of freedom it was difficult to
formulate Zerilli equation in terms of Newman-Penrose quantities.
   Although decoupled
equations do exist for $\Psi_2$ and $\Psi_{-2}$ \cite{Teukolsky, BardPres},
 even on a Kerr background, they are not
deformations\footnote{By deformation of the wave operator we mean a hyperbolic equation
such that when parameter $m$ vanishes it becomes the usual wave equation.}
of the usual wave equations in the asymptotic
region. Moreover, the NP special null tetrad (chosen in
\cite{Price} and \cite{BardPres}) is not symmetric with respect
to the interchange of null coordinates $u$ and $v$. In other
words it has to be chosen in a different way close to future
($\scri^+$) and past ($\scri^-$) null infinity. We would like to
convince the reader that the Teukolsky equation for
$\Psi_{-2}^{{\scriptscriptstyle\rm
Price}}=\Psi_4^{{\scriptscriptstyle\rm NP}}$ is not a primary
equation describing gravitational waves in asymptotic region (see
also \cite{Mino} for the review of the Teukolsky formalism). The
reasons are the following:
\begin{itemize}
\item The Teukolsky equation is not a deformation of a d'Alembert
equation in contrast to the Regge-Wheeler and Zerilli.
\item The initial data on a slice $t={\rm const.}$, instead of
position and momenta $({\bf x}, \dot{\bf x})$, corresponds rather
to second and third time derivatives of the position
$({\partial^2\over\partial t^2}{\bf x}, {\partial^3\over\partial t^3}{\bf x})$.
\end{itemize}
We may think about this equation as an evolution equation for the
acceleration $\ddot{\bf x}$. Usually, when we use Fourier
transform technique, it is not so important for plane waves which
variable (${\bf x}$ or $\ddot{\bf x}$) we are using as a
canonical position. However, choosing $\ddot{\bf x}$ we exclude
from the beginning all stationary solutions which are also
physically important for the wave operator.

In our gauge invariant quasilocal formalism we
  checked the ``peeling property'' for the
various components of the linearized Weyl tensor. In this paper we
calculated this property only for the axial degree of freedom
(governed by the Regge-Wheeler equation) and we showed that it is
valid in its original strong form
 similar to the case of a flat background (see the table in subsection
\ref{QNP}). However, for polar
 degree of freedom (governed by Zerilli equation
 \cite{Zerilli}, \cite{ISMC98}) we may have some
 obstructions. More precisely,
  because NP scalars are not gauge-invariant one
  could possibly ``damage'' their
asymptotics via gauge transformations. This is related to a
 more complicated behaviour of the asymptotic solutions
of the Zerilli equation and not obvious asymptotics of the gauge
transformations. Although we showed that for the axial part we
could not ``damage'' asymptotics via gauge transformation, it is
 not evident that the same is true for the polar degrees of
freedom. We shall elaborate upon this issue in a separate paper.

We believe that peeling phenomena for linearized gravitational field
 is a simpler property for the
gauge-invariants substituting NP scalars than for NP scalars
themselves.

We also hope that our results can be applied for improving
Christodoulou-Klainerman-Nicol\`o \cite{CKN} asymptotics on
$\scri^+$ for nonlinear Cauchy data ``sufficiently close'' to
Schwarzschild.

It is not easy to relate the results of \cite{F-L} with our
approach. The authors do not give explicit formulae for their
$\hat\Psi_k$. However, the deformation of
$\Psi_2^{\scriptscriptstyle\rm NP}$ proposed by them differs from
ours. In our case real and imaginary part of ``deformed''
$\Psi_2^{\rm NP}$ do not fulfill the same equation like in
\cite{F-L}. Moreover, the equations (3.77-3.81) on p. 847 suggest
that their invariants $\hat\Psi_2$, $\hat\Psi'_2$ are
quasilocally related to $\Psi_4^{\scriptscriptstyle\rm NP}$ and
$\Psi_0^{\scriptscriptstyle\rm NP}$. We would like to stress that
our invariants appear as the ``natural'' positions in the symplectic
analysis. The hamiltonian approach enables one to derive the
Regge-Wheeler equation, resp. the Zerilli equation, as an Euler-Lagrange second
order equation for the axial, respectively polar, part of the field
(see \cite{ISMC98}). This suggests that a deformed
$\Re\Psi_2^{\scriptscriptstyle\rm NP}$ should fulfill the Zerilli
equation but $\Im\Psi_2^{\scriptscriptstyle\rm NP}$ fulfills
Regge-Wheeler equation.

\appendix
\section{Linearization of Riemann tensor on vacuum background}
\label{lRt}
To fix the notation and for completeness
we present in this appendix the standard linearization formulae.
Let $M$ be a spacetime with
pseudoriemannian metric $g_{\mu\nu}$. We define
a linear perturbation of the metric  as
 \be
 h_{\mu \nu}\equiv \delta g_{\mu \nu} : =   g_{\mu \nu} -
 \eta_{\mu \nu}  \,.
 \ee
 where by $\eta_{\mu \nu}$ we denote the background metric.
 The tensor $h_{\mu\nu} $ is often called perturbation of the
metric $\eta_{\mu \nu}$. For the inverse metric the perturbation
has opposite sign:
\[ \delta g^{\mu \nu}  =  - h^{\mu \nu} := - \eta^{\mu \lambda}
\eta^{\nu\kappa} h_{\lambda\kappa} \, .\]

 Similarly, we have linearized Christoffel symbols (which are tensors):
 \be \label{dG} {\delta\Gamma^{\mu}}_{\nu\lambda}= \pl
\eta^{\mu\kappa}(h_{\kappa\nu ; \lambda}+h_{\kappa\lambda ;
\nu}- h_{\lambda\nu ; \kappa}) \, , \ee where as usual all
manipulations are with respect to the background metric,
background connection etc.
We have also linearized Riemann tensor:
 \be \label{dR} {\delta R^{\alpha}}_{\beta\mu\nu} =
{\delta \Gamma^{\alpha}}_{\beta\nu;\mu}-
{\delta\Gamma^{\alpha}}_{\beta\mu;\nu} \, . \ee
We would like to stress that in general for curved background
the linearization of the Riemann tensor depends on the
position of indices. Let us denote by
$
{r^{\alpha}}_{\beta\mu\nu} := {\delta R^{\alpha}}_{\beta\mu\nu}
$
the linearization of the Riemann tensor with one upper index, and
\[
r_{\kappa\beta\mu\nu} := {r^{\alpha}}_{\beta\mu\nu} \eta_{\alpha \kappa}
\, .
\]
Let us compare $r_{\kappa\beta\mu\nu}$ with
the corresponding linearization of the curvature tensor
with all indices lowered (i.e.
 $ \delta R_{\alpha \nu\rho\mu}
= \delta \left( g_{\alpha \kappa} R^\kappa
{_{\nu\rho\mu}}\right)$).
 The relation between $\delta R_{\alpha
\nu\rho\mu}$ and $r_{\alpha \nu\rho\mu}$ takes the form
 \be \label{zwiazek_r_R} \delta R_{\alpha \nu\rho\mu} :=
R_{\alpha \nu\rho\mu} ( g ) - {\cal R}_{\alpha \nu\rho\mu} ( \eta
) =  r_{\alpha \nu\rho\mu} + h_{\alpha \sigma} {\cal
R^\sigma}_{\nu \rho \mu}\,, \ee
where by  ${\cal
R}^\sigma{_{\nu \rho \mu}}$ we have denoted curvature tensor of the
background metric.

We would like to stress that tensor $r_{\alpha \nu\rho\mu}$
has not all symmetries of the usual curvature tensor.
This unpleasant property may be verified in the formulae below.\\
From definition of $r_{\mu\nu\sigma\rho}$ and equation (\ref{dG})
we can express (\ref{dR}) in terms of
$h_{\mu\nu}$ as follows
  \be \label{dr1}
 r_{\mu\nu\sigma\rho}= \frac12 \left(
h_{\mu\nu;\rho\sigma}-h_{\mu\nu;\sigma\rho}+h_{\mu\rho;\nu\sigma}-
h_{\nu\rho;\mu\sigma}-h_{\mu\sigma;\nu\rho}+h_{\nu\sigma;\mu\rho}
\right)
\, . \ee
The basic property of the curvature
 \be
h_{\nu\sigma;\mu\rho} - h_{\nu\sigma;\rho\mu} =
 h_{\alpha\sigma}{R^{\alpha}}_{\nu\mu\rho}+
h_{\nu\alpha}{R^{\alpha}}_{\sigma\mu\rho}
 \ee
 implies \be \label{dR2}
 2 r_{\mu\nu\sigma\rho} =
h_{\alpha\sigma}{\cal R^{\alpha}}_{\nu\mu\rho} + h_{\alpha\nu}
{\cal R^{\alpha}}_{\sigma\mu\rho} + h_{\mu\rho ; \nu\sigma}
-h_{\nu\rho ; \mu\sigma} - h_{\mu\sigma ; \nu\rho} +h_{\nu\sigma ;
\mu\rho} \, .
 \ee

 Let us denote by $r_{\nu\rho}$ a linearization of the Ricci tensor
 defined as follows
 \be\label{defr}
 r_{\nu \rho} := r^\sigma{_{\nu \sigma \rho}}=\eta^{\sigma \mu} r_{\mu \nu \sigma
 \rho}\, .
 \ee
If we assume that the background metric is Ricci flat
(i.e. is a solution of vacuum Einstein
equations)
\[ {\cal R_{\mu \nu}} = 0 \, , \]
 then we have the following form for the linearized Ricci tensor:
 \be \label{linear_ricci}
  r_{\nu\rho}= \frac12 \left(
  h_{\alpha\sigma}{\cal R^{\alpha}}_{\nu}{^\sigma}{_\rho}
  + {h^{\sigma}}_{\rho;\nu\sigma}+{h^{\sigma}}_{\nu;\rho\sigma}-
{{h_{\nu\rho}}^{;\sigma}}_{\sigma}-{h^{\sigma}}_{\sigma;\nu\rho}
\right)
\, .
 \ee

\section{Gauge invariance of ${ \mathbf y}$, ${ \mathbf y}_a$}
 \label{apendix_invariants}
 Let us begin with the gauge transformation (\ref{cech 1}) for
 $h_{aB}$:
\[ h_{aB} \rightarrow h_{aB}+\xi_{a,B}+\xi_{B,a}+ 2
\frac{\sk}{r} \varepsilon_a \xi_B \,. \]
and let us denote by $z_a := h_{aB||C} \varepsilon^{BC}$
a gauge-dependent axial part of $h_{aB}$. It is easy to check that
$z_a$ transforms as follows:
 \be\label{cza} z_a \rightarrow z_a + (\xi_{B||C}
\varepsilon^{BC} )_{,a}\,.
 \ee
 To compensate the gauge term
  $(\xi_{B||C} \varepsilon^{BC})_{,a}$ let us analyze the gauge
  transformation for the axial part of
 $\chi_{AB} = h_{AB}
-\pl \eta_{AB}  h_{CD} \eta^{CD}$:
\be\label{chiax}
 {{\chi_A}^B}_{||BC} \varepsilon ^{AC} \rightarrow
{{\chi_A}^B}_{||BC} \varepsilon ^{AC} + \eta^{EB} (\xi_{A||EBC}
+\xi_{E||ABC}) \varepsilon^{AC} \, .
\ee
Using the commutation relation for the second covariant
derivatives
\be\label{xiant} \xi_{A||EBC} = \xi_{A||ECB} + \xi_{F||E}
{\stackrel{_2}{R^ F}}_{ABC} + \xi_{A||F} {\stackrel{_2}{R^ F}}_{EBC} \,,\ee
where $\displaystyle {\stackrel{_2}{R^ F}}_{ABC}$ is the
curvature of $S^2$ given by (\ref{RnaS2}), from
(\ref{chiax}) and the identity $ \xi^E{_{||EAC}}
\ep^{AC} = 0 $, we obtain (\ref{cechowanie2})
\[ {{\chi_A}^B}_{||BC} \varepsilon ^{AC} \rightarrow
{{\chi_A}^B}_{||BC} \varepsilon ^{AC} + {{\xi_{A||C}}^E}{_E}
\varepsilon^{AC} + \frac{2}{r^2} \xi_{A||C} \varepsilon^{AC}\,, \]
 or in equivalent form
\be\label{cchi} {r^2}{{\chi_A}^B}_{||BC} \varepsilon ^{AC} \rightarrow
{r^2}{{\chi_A}^B}_{||BC} \varepsilon ^{AC} +
({\stackrel{\circ}{\Delta}+2}) (\xi_{A||C}
\varepsilon^{AC})\, . \ee Finally the gauge formulae
(\ref{cza}) and (\ref{cchi}) imply that
$ {\bf y}_a = (\stackrel{\circ}{\Delta}+2) z_a - ( r ^2
{{\xi_A}^B}_{||BC} \varepsilon ^{AC} )_{,a}$
is gauge independent.

Similarly, we can show gauge invariance of ${\bf y}$.
We shall derive the relation between
${\bf y}$  and the component $r_{abED} \ep^{ED} \ep^{ab}$
of the linearized Riemann tensor which is gauge independent
as has been shown in Section \ref{stR}.
From (\ref{dR2}) we obtain
\ber 2 r_{abED} & = &
 h_{aD \tr b || E} + h_{bE \tr a || D} - h_{aE \tr b || D}
 - h_{bD \tr a || E} +\nonumber \\
        & &  \skr \ep_b ( 2 h_{aD || E} -2 h_{a E || D} )
        + \skr \ep_a ( 2 h_{bE || D} -2 h_{b D || e} )\,.
\eer Moreover from $(\ep^{AB})_{\tr a} =
\frac {2 \sk}{r} \ep_a \ep^{AB}$ (cf. (\ref{1 poch ep_AB})) we get
 \ber 
 2 r_{abED} \ep^{ED} & = & (h_{aD
|| E} \ep^{ED})_{\tr b} + (h_{bE || D}\ep^{ED})_{\tr a} - (h_{aE
|| D} \ep^{ED})_{\tr b} - (h_{bD  || E}\ep^{ED})_{\tr a} \nonumber \\
& = & 2(h_{bE||D} \ep^{ED})_{\tr a} -2(h_{aE||D} \ep^{ED})_{\tr b}
\, . \eer and finally
\[
r_{abED} \ep^{ED} \ep^{ab} = 2(h_{bE||D} \ep^{ED})_{\tr a}
\ep^{ab}
\]
which implies the demanded relation for ${\bf y}$
\[
{\bf y} = \frac12 r^2 r_{abED} \ep^{ED} \ep^{ab} \, .
\]
This way we get gauge invariance of ${\bf y}$ from gauge independence
of $r_{abED}$. However,  equation (\ref{zwiazek}) also implies
this result from the gauge invariance of
 $ {\bf y}_{a}$.

\section{Compactification of Regge--Wheeler equation near $\scri^+$}
\label{RWs}

The equation (\ref{wave1}) in the coordinates $(u, \rho, x^A)$
can be rewritten\footnote{We remind that
$u:=  t - r - 2m \ln(r-2m)$, $\rho:=\frac1r$
and  $x^A$ are spherical angles.}  in the following form:
\be \label{wave2}
 2  \partial_u \partial_\rho \Phi +
\rho\partial_\rho \left[ k \partial_\rho (\rho\Phi) \right] + \dtwo
\Phi + 8m \rho \Phi =0  \ee
where $ \Phi: = \rho^{-1}{\mathbf y}$ and $k=1-2m\rho$.
Standard arguments, using domain of dependence considerations
together with conformal covariance of Equation (\ref{wave2}),
 show that smooth initial data which are compactly
supported on some Cauchy hypersurface for the
Kruskal--Schwarzschild spacetime lead to the solutions of
equation (\ref{wave2}) such that the rescaled $\Phi$ smoothly
extends across $\scri^+$ (cf. \cite{FGF2}).
This means that the assumption (\ref{szereg1}) is fulfilled
for a large class of solutions for the Regge-Wheeler equation (\ref{wave1}).
On the other hand,
no conditions on initial data which are not compactly supported
are known, which would guarantee smoothness of solutions across
$\scri^+$.

\section{Axial part of linearized Einstein equations}
\label{apendix_rownania}
 The equation (\ref{r2}) contains the component
 $\stackrel{\circ}r{^A}{_{B||AC}} \ep^{BC}$
where by $\stackrel{\circ}{r}_{AB}\equiv TS(r_{AB})$
we denote traceless symmetric part of tensor $r_{AB}$ (cf. (\ref{btr}))
\[
 \stackrel{\circ}{r}_{AB} := r_{AB} - \pl \eta_{AB} r_{CD} \eta^{CD}
 \equiv TS(r_{AB}) \, , \quad TS(h_{AB}) \equiv \chi_{AB} \, .
\]
 The general formula (\ref{linear_ricci})
 takes the following form for $r_{AB}$ on Schwarzschild
 background:
 \be \label{r_ab}
  2r_{AB} =  h_{ab}C^{a}{_{A}}{^b}{_B}
  + h_{CD}C^{C}{_{A}}{^D}{_B}
  + {h^a}_{B;Aa} + {h^F}_{B;AF} +{h^a}_{A;Ba} +{h^F}_{A;BF}
  \ee
  \[
  +{{h_{AB;}}^a}_a -
{{h_{AB;}}^F}_F - h_{;AB}
\]
where $h:=\eta^{\mu\nu}h_{\mu\nu}$.

Let us notice that the first two terms in (\ref{r_ab})
 containing
background Riemann $C_{\mu\nu\lambda\kappa}$ are proportional to the metric
 $\eta_{AB}$ so they disappear in  $\stackrel{\circ}{r}_{AB}$.
 To give explicit expression for $r_{AB}$ in terms of $h_{\mu\nu}$
we need first to examine the terms $ h_{aB;Ab}$, $h_{AB;CD}$ and
$h_{AB;ab}$.
 Using extensively the formulae from Section \ref{lfS}
 one can show that the following equalities hold:
 \ber
h_{bB;Ad} & = & h_{bB||A \tr d} +
\skr ( h_{AB \tr d} \ep_b +2 h_{bB||A} \ep_d
-\ep^f h_{bf \tr d} \eta_{AB} ) + \nonumber \\
          &  &\hspace*{-1ex} \frac k {r^2}
          (3 h_{AB} \ep_d \ep_b + h_{bf} \ep^f \ep_d \eta_{AB})
          + \frac m {r^3} ( h_{bd} \eta_{AB} - h_{AB} \ep_b \ep_d
          ) \label{h1}
 \eer
 \ber
  h_{AB;CD} & = & h_{AB||CD} - \skr \ep^a( h_{aB||D} \eta_{AC}
 + h_{Aa||D} \eta_{BC} + h_{aB||C} \eta_{AD} + \nonumber \\ & &
 h_{Aa||C}\eta_{BD} + h_{AB \tr a} \eta_{CD} ) +   \frac{k}{r^2} (
 h_{af} \ep^a
 \varepsilon^f \eta_{AD} \eta_{BC} + h_{af}\ep^a \varepsilon^f \eta_{BD}
 \eta_{AC} + \nonumber \\ & &
 - h_{CB} \eta_{AD} - h_{AC} \eta_{BD} - 2 h_{AB} \eta_{CD} ) \label{h2}
 \eer
 \ber   h_{AB;ab} & = & h_{AB \tr ab} + 2\skr (
h_{AB \tr b} \varepsilon_a +  h_{AB \tr a} \varepsilon_b ) + 6
\frac {k}{r^2} h_{AB} \varepsilon_b \varepsilon_a
 - \frac{2m}{r^3}
\eta_{ab}  h_{AB} \label{h3}
 \eer
 To simplify the analysis we assume that $h_{AB}=0$.
 The final gauge-invariant result is not dependent on
 this assumption but we shall see
 that formulae are much simpler when we assume this gauge condition.
Moreover, we may neglect all terms proportional to the metric
  $\eta_{AB}$ because they drop out when we pass to the traceless
  part.
From (\ref{h1}--\ref{h3}) we obtain
\[ {h^a}_{B;Aa} \backsimeq
{h^a}_{B||A \tr a} + 2 \skr h_{aB||A} \varepsilon^a \]
\[ {h^F}_{B;AF} \backsimeq
 - 4 \skr h_{aB||A} \varepsilon^a \]
\[ {h^a}_{A;Ba} \backsimeq
{h^a}_{A||B \tr a} + 2 \skr h_{aA||B} \varepsilon^a\]
\[ {h^F}_{A;BF} \backsimeq
 - 4 \skr h_{aA||B} \varepsilon^a\]
\[ {{h_{AB;}}^a}_a \backsimeq
0 \]
\[ {{h_{AB;}}^F}_F \backsimeq
 - \skr \varepsilon^a ( 2h_{aB||A} + 2h_{aA||B})
\]
\[ h_{;AB} \backsimeq
h_{||AB} \]
where $\backsimeq$ denotes an equality modulo trace:
\[ t_{AB} \backsimeq \tau_{AB} \iff TS(t_{AB}) = TS(\tau_{AB})\, . \]
 The above formulae enables one to rewrite (\ref{r_ab}) in a simpler
 form:
\be \label{arAB}
2r_{AB} \backsimeq {h^a}_{B||A \tr a} + {h^a}_{A||B \tr a} - h_{||AB}\, .
\ee
One can easily check the following identity on $S^2$:
\[ \varepsilon^{AC}
\left( h_{||AB} -\frac12\eta_{AB} h^{||D}{_D}\right){^{||B}}_C
    = 0 \, ,\]
so the last term in (\ref{arAB}) drops out when we pass to
${\stackrel{\circ}{r}_A{^B}}_{||BC} \varepsilon^{AC}$. Hence
\ber
2{\stackrel{\circ}{r}_A{^B}}_{||BC} \varepsilon^{AC} & = &
\left(
{{{h^a}}_{B||A}}_{\tr a} +
{{{h^a}}_{A||B}}_{\tr a} -\eta_{AB} \eta^{DE} {{{h^a}}_{D||E}}_{\tr a}
\right){^{||B}}_C \varepsilon^{AC} \nonumber \\&=&
\varepsilon^{AC}\eta^{FB} \left(
{{{h^a}}_{B||A}}_{\tr a} +
{{{h^a}}_{A||B}}_{\tr a} -\eta_{AB} \eta^{DE} {{{h^a}}_{D||E}}_{\tr a}
\right)_{||FC} \, .
 \eer
Taking into account that
 \[ {\eta^{AB}}_{,a} = 2r ( \varepsilon_a \sqrt{k}) \eta^{AB}\, ,
 \quad
 {\varepsilon^{AB}}_{,a} = 2r ( \varepsilon_a \sk ) \varepsilon^{AB} \]
or simply $(r^2{\eta^{AB}})_{,a}=(r^2{\varepsilon^{AB}})_{,a}=0$,
we get
 \ber
 2r^4{{\stackrel{\circ}{r}_A}^B}_{||BC} \varepsilon^{AC} & = &
\left[ r^2\varepsilon^{AC}r^2\eta^{FB} \left(
{h^a}{_{B||A}} +
{{{h^a}}_{A||B}} -\eta_{AB} \eta^{DE} {{{h^a}}_{D||E}}
\right)_{||FC} \right]_{\tr a}
 \nonumber \\
& = & \left[ r^2 (\dtwo +2){h^a}{_{A||B}}
\varepsilon^{AB} \right]_{\tr a} \, .\eer
Moreover, the gauge condition
 ${{\chi_A}^B}_{||BC} \varepsilon ^{AC}=0$
 implies that ${\bf y}_a=(\stackrel{\circ}{\Delta}+2)
 h_{aA||C} \varepsilon^{AC}$
 and we obtain the demanded result (\ref{r2}), namely
 \begin{equation} 2 r^4 { \stackrel{\circ}{r} }_A{^B}_{||BC}
\varepsilon^{AC} =  \left( {r^2}{\bf y}^a \right)_{\tr a} \, .
\end{equation}

The equation (\ref{ryd}) can be derived in a similar way.
Let us start with
\begin{equation} \label{2_1}
r_{Bd}  = r_{adcB} \eta^{ac} + r_{AdCB}
\eta^{AC} = \delta R_{adeB}\eta^{ae}+ \delta R_{ABEd}\eta^{AE}
-\frac{m}{r^3} h_{dB}
\end{equation}
where the corresponding components of the linearized Riemann tensor
$\delta R_{\mu\nu\lambda\kappa}$ have the following explicit form
\cite{adamj}:
\ber
2 \delta R_{adeB} & = &  h_{aB \tr de} - h_{ae \tr d||B} - h_{dB\tr ae}
+ h_{de \tr a || B} +  \skr ( h_{de||B} \ep_a - h_{ae||B} \ep_d )
\nonumber \\
         &  & + \skr ( h_{aB \tr e} \ep_d - h_{dB \tr e} \ep_a
         + h_{Be \tr a} \ep_d - h_{Be \tr d} \ep_a ) \, ,
         \eer
\ber
2 \delta R_{ABEd} & = & h_{Ad||BE} - h_{Bd||AE} + h_{BE||A \tr d}
 - h_{AE||B \tr d} +  2 \frac k {r^2} \ep^a \ep_d ( h_{Aa} \eta_{EB}
  -  h_{Ba} \eta_{EA})\nonumber \\
         &  &+ \skr ( 2 h_{BE||A} \ep_d - 2 h_{AE||B} \ep_d
         + h_{ad||A} \ep^a \eta_{BE}
         - h_{ad||B} \ep^a \eta_{AE})  \nonumber \\
         &  &+ \skr \ep^a ( h_{Bd \tr a}  \eta_{AE}
         - h_{Ad \tr a}  \eta_{BE} + h_{Aa \tr d} \eta_{EB}
         - h_{Ba \tr d} \eta_{EA})   \,.
 \eer
Hence from (\ref{2_1}) we obtain
\ber 2r_{Bd} & = & h^a{_{B\tr da}} - h_{dB\tr a}{^a}
- h_{dB||A}{^A} +2\skr ( h^a{_{B\tr a}}\ep_d - h_{aB\tr d}\ep^a )+
\frac k {r^2} (h_{dB} - 2 \ep^a \ep_d h_{aB}) \nonumber
\\ & & + \left( h_{BA}{^{||A}} - H_{||B} \right)_{\tr d}
+\left[ h_d{^a}{_{\tr a}} - h_d{^a}{_{\tr a}} - \skr h_a{^a}\ep_d +
 h^A{_{d||A}} \right]_{||B} \, . \label{rBdh}
 \eer
 Let us notice that the all terms in square brackets  (gradient)
in the above equality do not contribute
to axial part $r_{Bd||D} \ep^{BD}$. Moreover, if we
assume that
$h_{AB}=0$ (as a gauge condition) we conclude
that the ``second line'' in equation ({\ref{rBdh}) drops out
and the rest gives
\begin{equation} \label{tsr}
2r_{Bd||D} \ep^{BD} = - h_{dB}{^{||A}}{_{AD}}\ep^{BD}
+ z^a{_{\tr da}} - z_{d \tr a}{^{a}} + 4\skr\ep^a
 (z_{d\tr a} - z_{a\tr d}) - \frac{1}{r^2} z_d
\end{equation}
where $z_a=h_{aA||C} \varepsilon^{AC}$ (cf. \ref{cza}))
and we used the following identity:
\[ (\ep^{BD} h^a{_B})_{\tr da} - (\ep^{BD} h_{dB})^{\tr a}{_a} +
   4\skr\ep^a \left[ (\ep^{BD} h_{dB})_{\tr a} - (\ep^{BD} h_{aB})_{\tr d}
   \right] = \]
\[ =\ep^{BD} \left[ h^a{_{B\tr da}} - h_{dB\tr a}{^a}
  +2\skr ( h^a{_{B\tr a}}\ep_d - h_{aB\tr d}\ep^a )
 +2\frac k{r^2} (h_{dB} - \ep^a \ep_d h_{aB}) + 2\frac m{r^3} h_{dB}
 \right] \]
 implied by (\ref{2 poch ep_AB}). Using one more identity
 \[ h_{dB}{^{||A}}{_{AD}}\ep^{BD} = \frac1{r^2} (\dtwo +1)
 h_{dB||D}\ep^{BD} \]
 which is a straightforward consequence of the commutation relation
 (\ref{xiant})
 for covariant derivatives on $S^2$, we get
 \ber
2r_{Bd||D} \ep^{BD} & = & (z_{a \tr d} - z_{d \tr a}){^{\tr a}}
- \frac {4\sk}{r}  \ep^a (z_{a \tr d} - z_{d \tr a})
 -\frac1{r^2} (\dtwo +2) z_d
\eer
 or in equivalent form
\begin{eqnarray}
-2r^4 r_{Ba||D} \ep^{BD}  & = &
\left[ r^4 ( z_{a\tr b} - z_{b \tr a} )\right]^{\tr b} + r^2 {\mathbf y}_a
\, .
\end{eqnarray}
The identity
\be
 \label{dod1} z_{a\tr d} - z_{d \tr a} = \ep_{ad}
\ep^{bc} z_{b \tr c}  = \ep_{ad} \frac{{\mathbf y}}{r^2}
 \ee
similar to (\ref{zwiazek}) implies
\begin{equation}\label{rybc}
-2r^4 r_{Bd||D} \ep^{BD}
=( r^2 {\mathbf y} )^{\tr b} \ep_{ab} + r^2 {\mathbf y}_a
\end{equation}
which finally gives equation (\ref{ryd}) even if we relax the gauge
condition
$h_{AB}=0$ because both sides of (\ref{rybc}) are gauge independent.

\section{Multipoles and traceless tensors}\label{multipole}

Let $P^k$ denotes the space of polynomials of degree $\leq k$ in
${\mathbb R}^3$. If function $f$ is defined in a neighbourhood of a
unit sphere $S^2$, we
can denote by $Rf$ its restriction to $S^2$.\\
In \cite{erdelyi} one can find well-known theorem that $R(P^k)$ is the
direct sum $\sum_{l=0}^{k} SH^l$, where $SH^l$ denotes the space of
spherical harmonics of degree $l$, ($g\in SH^l \Longleftrightarrow
 {\dtwo}g= -l(l+1)g$).

Let $t$ be a tensor field in a neighbourhood of $S^2$ in ${\mathbb R}^3$
and by $TS(Rt)$ we denote traceless symmetric part of $Rt$ on $S^2$.
The following identity holds:
\begin{equation}\label{TS} TS\left(R(t_{A_1\ldots A_k|B})\right) =
TS\left( (Rt_{A_1\ldots A_k})_{||B}\right) \end{equation} where ``$|$''
denotes covariant derivative with respect to the flat 3-metric on
${\mathbb R}^3$ and  ``$||$'' stands for covariant derivative on
$S^2$. To prove (\ref{TS}) we can observe the following:
\[ t_{A_1\ldots A_k|B}=t_{A_1\ldots A_k||B} +
\frac 1r \eta_{A_1 B}t_{3 A_2\ldots A_k} + \frac 1r \eta_{A_2 B}t_{A_1
3 \ldots A_k} + \ldots + \frac 1r \eta_{A_k B}t_{A_1 A_2\ldots A_{k-1}
3}
\]
and $TS\left(\eta_{AB} t_{A_1\ldots A_n}\right)=0$.

We use spherical coordinates in ${\mathbb R}^3$:

$x^3:=r$, $(x^A),(A=1,2)$, $(x^1=\theta, x^2=\phi)$

$\eta_{AB}=r^{2}\kolo\gamma_{AB}$, $\eta_{33}=1$

$\eta^{AB}=r^{-2}\kolo\gamma^{AB}$, $\eta^{33}=1$, $\eta^{aA}=0$

$\Gamma^3{_{AB}}=-\frac 1r \eta_{AB}$, $\Gamma^A{_{3B}}
 = \frac 1r \delta^A{_B}$\\[4ex]
{\em Theorem:}
\[ p\in P^{n-1} \Longrightarrow TS\left(Rp_{||A_1\ldots A_n}\right)=0 \]
{\em Proof:}\\
Let us denote by $(x^k)$ cartesian coordinates in ${\mathbb R}^3$ such
that $S^2$ corresponds to the surface: $(x^1)^2+(x^2)^2+(x^3)^2=1$.
Transformation rule for transition from spherical to Cartesian coordinates gives as
following
\[ p_{|A_1\ldots A_n}=x^{i_1}_{,A_1} x^{i_2}_{,A_2} \cdots x^{i_n}_{,A_n}
p_{|i_1\ldots i_n} = x^{i_1}_{,A_1} x^{i_2}_{,A_2} \cdots x^{i_n}_{,A_n}
p_{,i_1\ldots i_n} \, , \] and $p_{,i_1\ldots i_n}=0$ because degree of the
polynomial $p$ is not greater than $n-1$. From (\ref{TS}) and
$p_{|A_1\ldots A_n}=0$
 we get the result.$\Box$\\
 In particular for $n=2$ we can easily obtain that $\chi^{AB}{_{||AB}}$ is
orthogonal to the space $SH^0\oplus SH^1$. More precisely, if $f\in
SH^0\oplus SH^1$ then $TS\left( f_{||AB}\right)=0$ and
\[ 0=\int_{S^2}\chi^{AB}TS\left( f_{||AB}\right)
 =\int_{S^2} \chi^{AB}{_{||AB}} f \]

Let us consider the following diagram:
\[
\begin{array}{ccccccccc}
V^0\oplus V^0 & \stackrel{i_{01}}{\longrightarrow} & V^1 &
\stackrel{i_{12}}{\longrightarrow} & V^2 &
\stackrel{i_{21}}{\longrightarrow} & V^1
& \stackrel{i_{10}}{\longrightarrow} & V^0 \oplus V^0 \\
\Big\downarrow\vcenter{\rlap{$\scriptstyle Fl$}} &  &
\Big\downarrow\vcenter{\rlap{$\hat{} $}} &  &
\Big\downarrow\vcenter{\rlap{$\hat{} $}} &  &
\Big\downarrow\vcenter{\rlap{$\hat{} $}} &  &
\Big\downarrow\vcenter{\rlap{$\scriptstyle Fl$}}  \\
V^0\oplus V^0 & \stackrel{i_{01}}{\longrightarrow} & V^1 &
\stackrel{i_{12}}{\longrightarrow} & V^2 &
\stackrel{i_{21}}{\longrightarrow} & V^1 &
\stackrel{i_{10}}{\longrightarrow} & V^0 \oplus V^0
\end{array}
\]
where the mappings and the spaces are defined as follows:
\[ i_{01}(f,g)=f_{||A}+\varepsilon_A{^B}g_{||B} \]
\[ i_{12}(v)= v_{A||B}+ v_{B||A}-\kolo\gamma_{AB}v^C_{||C} \]
\[ i_{21}(\chi)= \chi_A{^B}{_{||B}} \]
\[ i_{10}(v)=\left( v^A_{||A}, \varepsilon^{AB}v_{A||B} \right) \]
\[ Fl(f,g)=(g,f) \quad {\hat v}_A=\varepsilon_A{^B}v_{B} \quad
 {\hat\chi}_{AB}=\varepsilon_A{^C}\chi_{CB} \]
 \noindent
$V^0$ -- scalars on $S^2$ \\
 $V^1$ -- covectors on $S^2$\\
$V^2$ --symmetric traceless tensors on $S^2$.

 We have denoted by $\dtwo$ the laplacian on $S^2$.
The following equality
\[ i_{10} \circ i_{21} \circ i_{12} \circ i_{01} = {\dtwo}({\dtwo}+2) \]
shows that if we restrict ourselves to the spaces $\overline V^0:=V^0
\ominus [SH^0\oplus SH^1 ]$
 (${\dtwo}({\dtwo}+2)\overline V^0=\overline V^0$) and
 $\overline V^1=V^1\ominus[i_{01}( SH^1)]$
  ($({\dtwo}+1)\overline V^1=\overline V^1$) then the all mappings in
the above diagram become isomorphisms.\\
We define {\em mono-dipole-free scalar} as an element of $\overline
V^0$, {\em mono-dipole-free covector} belongs to $\overline V^1$ and
any symmetric traceless tensor on $S^2$ is {\em mono-dipole-free}.


\end{document}